\documentstyle[12pt,epsf]{article}
\epsfverbosetrue
\def\figlabel#1{\xdef#1{\thefigure}}
\def\fig#1{fig.~#1}

\def\figalign#1#2#3#4#5#6{
\begin{figure}
\centerline{
\hbox to 2.5truein{\vtop{\hsize=2.5truein\epsfxsize=6cm
\centerline{\epsfbox{#1} }
\caption[]{#3}
\figlabel{#2}
}}
\qquad\hbox to 2.5truein{\vtop{\hsize=2.5truein\epsfxsize=6cm
\centerline{\epsfbox{#4} }
\caption[]{#6}
\figlabel{#5}
}}
}
\end{figure}
}

\def\be{\begin{equation}}
\def\ee{\end{equation}}
\def\bea{\begin{eqnarray}}
\def\eea{\end{eqnarray}}

\def\b{\beta}

\def\d{\delta}
\def\pa{\partial}

\begin{document}
\begin{titlepage}
\begin{flushright}
{ ~}\vskip -1in
CERN-TH/97-144\\
US-FT-21/97, UB-ECM-PF 97/10\\
hep-th/9707017\\
June 1997\\
\end{flushright}
\vspace*{20pt}
\bigskip
\begin{center}
 {\Large SOFTLY BROKEN $N=2$ QCD \\
   \bigskip
\Large WITH MASSIVE QUARK HYPERMULTIPLETS, II}
\vskip 0.9truecm

{Luis \'Alvarez-Gaum\'e$^{a}$,
Marcos Mari\~no$^{b}$ and 
Frederic Zamora$^{a,c}$.}

\vspace{1pc}

{\em $^a$ Theory Division, CERN,\\
 1211 Geneva 23, Switzerland.\\
 \bigskip
  $^b$ Departamento de F\'\i sica de
Part\'\i culas,\\ Universidade de Santiago
de Compostela,\\ E-15706 Santiago de Compostela, Spain.\\
  \bigskip
$^c$ Departament d'Estructura i Constituents de la Materia,
\\ Facultat de F\'\i sica, Universitat de Barcelona,\\ 
Diagonal 647, E-08028 Barcelona, Spain.}\\

\vspace{5pc}

{\large \bf Abstract}
\end{center}
We analyze the vacuum structure of $N=2$, $SU(2)$ QCD 
with massive quark hypermultiplets, once supersymmetry is 
soflty broken down to $N=0$ with dilaton and mass spurions. 
We give general expressions for the low energy couplings 
of the effective potential in terms of elliptic functions
to have a complete numerical control of the model.
We study in detail the possible phases of the theories 
with $N_f=1,2$ flavors for different values of the 
bare quark masses and the supersymmetry breaking parameters
and we find a rich structure of first order phase transitions. 
The chiral symmetry breaking pattern of the $N_f=2$ theory is considered, 
and we obtain the pion Lagrangian for this model up to two derivatives. Exact 
expressions are given for the pion mass and $F_{\pi}$ in terms 
of the magnetic monopole description of chiral symmetry breaking.


\end{titlepage}

\def\theequation{\thesection.\arabic{equation}}

\section{Introduction}
\setcounter{equation}{0}

This is the second part of a series of papers where we study the 
soft breaking of $N=2$ QCD with massive matter.
In the first paper of the series \cite{amz} (referred as (I) from now on)
we focused on two well differentiated but necessary goals. The first one
was to have complete numerical control of 
the exact Seiberg-Witten solution for $SU(2)$ gauge group and $N_f 
\leq 3$ massive quark hypermultiplets. We presented a general framework to 
obtain explicit expresions for  the Seiberg-Witten periods $(a_D(u), a(u))$
 and all the effective couplings of the nonsupersymmetric low energy 
effective Lagrangian at {\it any} point of the $u$-plane. This powerful 
method was based on the use of uniformization theory for the 
Seiberg-Witten elliptic curves. 

 The second goal of the first paper was to study the logic of the 
general soft breaking of massive $N=2$ QCD down to $N=0$ by the introduction
of $N=2$ spurion fields. We promoted the masses in  the bare 
Lagrangian to $N=2$ vector superfields by gauging the baryon 
numbers of the quark hypermultiplets. We obtained an effective 
$SU(2) \times U(1)^{N_f}$ Seiberg-Witten model, where the dynamics of 
the $N_f$ $U(1)$ baryon gauge symmetries is frozen and we turn on the 
auxiliary fields of the corresponding $N=2$ vector superfields to get 
additional supersymmetry breaking parameters. 
We also embedded the effective $SU(2) \times U(1)^{N_f}$
Seiberg-Witten model into a pure gauge theory with higher rank group,
where the additional degrees of freedom with ``magnetic'' baryon numbers 
different from zero, became infinitely 
heavy and decoupled in a limiting region of the moduli space.
We analysed in full detail all the monodromy properties and the internal 
consistency of these softly broken models. Finally, we gave the general
expression of the exact effective potential of the softly broken theory 
when both dilaton and mass spurions are included.

In this second paper, we perform a complete analysis of the 
nonperturbative vacuum structure and the large distance physics of
the softly broken $N=2$ massive QCD. The structure of the paper 
is the following:

In section 2 we present the $N=2$ soft breaking terms of the 
bare Lagrangian and discuss their main physical features.
For completeness, we also give the final formulae for $(a_D(u), a(u))$,
the dual mass $m_D$, and all the effective couplings, $\tau^{AB}$,
in terms of elementary elliptic functions. These explicit formulae can
be implemented numerically in the Mathematica program. 
This makes possible
the extraction of the nonperturbative effects encoded in the 
nonsupersymmetric effective 
potential and to unveil the rich phase structure of these models. 
At the end of the section,
we give an exact expresion of the squark condensates $<{\tilde q}_f q_f>$, 
the ``electric'' order parameters of chiral symmetry breaking,
in terms of the low energy couplings and the monopole condensate.

In section 3, we focus on the vacuum structure with one massive
quark hypermultiplet. The phase structure of the theory depends on the
values of the bare mass $m$, and the two supersymmetry breaking parameters: 
$f_0$ for the dilaton spurion and $f_m$ for the mass spurion. The phase
diagram is divided in two well defined regions. When $m < m_c$, 
where $m_c$ is the critical mass for the
Argyres-Douglas point to occur \cite{apsw}, the theory is in different
Higgs-confining phases. There are first order phase
transitions among these phases associated to the condensation 
of ``bound states'' made of mutually non-local degrees of freedom, with
total magnetic charge $n_m=2$ or $n_m=3$. For $m>m_c$,
when only the dilaton spurion is turned on, 
 
the vacuum structure flows to the
pure $SU(2)$ one analysed in \cite{soft} an expected result 
because of the quark decoupling under 
the flow of the renormalization group. The theory is then 
in a ``magnetic'', confining phase. 
But when supersymmetry is broken with the mass spurion 
$f_m$, the ground state is associated
 to an elementary quark condensate.
This purely ``electric'' vacuum goes to infinity in the $u$-plane
as the mass is increased, and in the decoupling limit we obtain 
a flat potential.  
The moduli space of vacua of $N=2$ supersymmetric Yang-Mills theory
\cite{swone} is then recovered.

In the rest of the paper we focus on the nonperturbative physics at large 
distance of the $N_f=2$ theory with small bare masses. In this case we have a 
QCD-like situation, where the number of light flavors is equal to the
number of colors. For massless quark hypermultiplets, the flavor symmetry
group is $SU(2)_- \times SU(2)_+$. At low energy, the relevant degrees 
of freedom are light magnetic BPS states in flavor group 
representations $({\bf 2}, {\bf 1})$ or $({\bf 1}, {\bf 2})$.
Chiral symmetry breaking is then driven by the 
condensation of these light monopoles \cite{swtwo}.

In section 4 we present the phase structure of the $N_f=2$ softly
broken theory, as a function of the different values of the dilaton and
mass spurions. For supersymmetry breaking dominated by the dilaton 
spurion, the vacuum is in the magnetic region 
$u \sim \Lambda^2_2 /8$, with first
order phase transitions between mutually local minima as the values of the 
mass spurions are slightly changed. The pattern
of chiral symmetry breaking is $SU(2)_- \times SU(2)_+ \rightarrow 
SU(2)_+$. If the supersymmetry breaking is dominated by the mass spurions,
there is a new kind of first order phase transition between mutually
non-local minima. The minimum can jump from the monopole 
to the dyon region $u \sim -\Lambda^2_2 /8$,
and in this way the opposite pattern of chiral symmetry breaking 
$SU(2)_- \times SU(2)_+ \rightarrow SU(2)_-$ is realized. Finally, we show that 
there are critical values of the bare masses such that a new minima is 
generated at $u=0$, due to the condensation of a ``bound state''
in the $({\bf 2}, {\bf 1}) \bigoplus ({\bf 1}, {\bf 2})$ representation of 
the flavor group. This could be an indication of a new phase where 
the pattern of chiral symmetry breaking is 
$SU(2)_- \times SU(2)_+ \rightarrow SU(2)_V$.

Finally, in section 5 we give the pion Lagrangian
up to two derivatives, for the chiral symmetry breaking 
pattern 
$SU(2)_- \times SU(2)_+ \rightarrow SU(2)_+$. We obtain expresions 
for the pion 
decay constant $F_\pi$ and the pion masses $M_\pi$ in terms of the 
bare quark masses and the supersymmetry breaking parameters. 
This gives a connection between these phenomenological parameters and the 
magnetic monopole description of chiral symmetry breaking. We think that, 
although 
the model we study has many obvious differences with respect to ordinary QCD, 
some 
features of this connection can be present in more realistic models. 
The pion 
Lagrangian parameters we obtain are  intrinsically nonperturbative, 
and they have a natural expansion in terms of 
inverse powers of the dynamically generated scale $\Lambda$.

\section{Soft breaking of $N=2$ QCD with Massive Quark Hypermultiplets }

\setcounter{equation}{0}

\subsection{Breaking supersymmetry. The Bare Lagrangian}

To break $N=2$ supersymmetry down to $N=0$ in the $SU(2)$ theory with 
$N_f$ flavors , we promote the scale 
$\Lambda_{N_f}$ \cite{soft, softdos} and the masses $m_f$ \cite{amz} 
to $N=2$ vector superfields, and then we 
freeze the scalar and auxiliary components to be constants. 
The dilaton spurion 
is introduced through the relation $\Lambda_{N_f} = \mu_0{\rm e}^{iS}$,
with $\mu_0$ the UV cut-off scale of the bare Lagrangian. 
The RG equation 
for the $SU(2)$ theory with $N_f$ hypermutliplets gives the relation  
$\Lambda_{N_f}^{4-N_f} \sim {\rm e}^{i \tau}$, where we use the normalization 
of \cite{swtwo} for the coupling constant, $\tau = 8 \pi i/g^2 + \theta /\pi$.
 We 
then obtain $s = \pi \tau /(4-N_f)$, and the spurion superfield $S$ 
appears in the 
classical prepotential as 
\be
{\cal F} = {4-N_f \over 2 \pi}S A^2.
\label{clas}
\ee
The bare Lagrangian reads, once the auxiliary fields 
of the dynamical superfields are integrated out,
$$ 
{\cal L} = {\cal L}_{N=2} + (4-N_f) \Bigg[ {\rm Im} \Big( {F_0 \over 8 \pi ^2}
\lambda \lambda + {{\overline F}_0 \over 8 \pi ^2}\psi \psi \Big) + 
{{\sqrt 2} D_0 \over 8 \pi^2}{\rm Im}\Big( i \lambda \psi \Big) \Bigg] 
$$
$$
-{(4-N_f) D_0 
\over \pi {\rm Im \tau}}f_{abc} ({\rm Im} \phi^a)\phi^{b} {\bar \phi}^c  
- 
{(4-N_f)  
\over \pi {\rm Im }\tau }\Big(F_0{\tilde q}_f ({\rm Im} \phi) q_f 
+ {\rm h.c.} \Big)
$$
$$
- {(4-N_f) D_0 
\over \pi {\rm Im \tau}}\Big({ q}^{\dagger}_f ({\rm Im} \phi) q_f-
{\tilde q}_f ({\rm Im} \phi){\tilde  q}^{\dagger}_f  \Big)
$$
$$
- {(4-N_f)^2 \over 4 \pi^3 {\rm Im \tau}}
\Big({1 \over 2} D_0^2 + |F_0|^2 \Big) ({\rm Im} \phi )^2 
$$
\be 
+{\sqrt 2} S_f \Big(F_f {\tilde q}_f q_f + {\rm h.c.} \Big)+ 
S_f D_f (|q_f|^2-|{\tilde q}_f|^2).
\label{lagrangiano}
\ee
In this Lagrangian, $\lambda$, $\psi$ are the gluinos, $\phi$ 
is the scalar component of the $N=2$ vector superfield, and $q_f$, 
${\tilde q}_f$ are the squarks.
The constants $S_f$ are the baryon numbers of the quark hypermultiplets. 
In the numerical study of the model, we set them equal 
to one\footnote{For $S_f \not= 1$, the Seiberg-Witten elliptic curves 
given in \cite{swtwo} must be properly modified.}.
As the prepotential has an analytic dependence on the spurion superfields, 
the effective Lagrangian 
up to two derivatives and four fermions terms for the $N=0$ theory 
described by (\ref{lagrangiano}) is 
given by the exact Seiberg-Witten solution once the spurion superfields are 
taken into account. This gives the exact effective potential at leading order 
and the vacuum structure can be determined. 

Notice that the terms involving mass spurions can be regarded as 
additional mass terms for the squarks, but not all of them are 
positive definite. More precisely, the graded trace of the 
mass matrix still vanishes after 
this soft breaking of supersymmetry, and the additional mass terms for the 
squarks are grouped in pairs with opposite signs. This is apparent in the 
$D_f$ terms in (\ref{lagrangiano}). We then expect a vacuum structure 
similar to the one found in \cite{mimura} for the soft breaking of $N=1$ 
supersymmetry, although here the structure is more complicated due 
to the presence of the Higgs field $\phi$. These kinds of terms have two 
consequences for the vacuum structure of the theory: first, for certain values 
of the spurions and the masses the resulting potential can be unstable, as 
the vacuum energy can be unbounded from below. Second, these terms favour 
an squark condensate in a Higgs phase. 
Both issues were discused in \cite{mimura} in the context of $N=1$ 
supersymmetry, and here we will adress them using the exact 
effective potential of the theory.

In general, the bare Lagrangian (\ref{lagrangiano}) will not be CP invariant,
since $\tau, m_f, F^A$ are arbitrary complex parameters. 
There are some particular situations where we still can have CP invariance.
We can asign to the bare masses $m_f$ and the spurion dilaton $F_0$ 
an $R$-charge two. If these parameters have the same complex phase,       
we can perform an anomalous $U(1)_R$ transformation such that for
 a value of the bare $\theta$ angle,
CP is a symmetry of the bare Lagrangian.
The other situation where CP is not lost is when $\theta$ is equal zero, 
the bare quark masses $m_f$
 are real and positive and all supersymmetry breaking 
parameters $F^A$, $A=0,1,..., N_f$, have the same complex phase. In this case,
 we can make a non anomalous 
transformation on the phases of the squarks and the gluinos 
such that the common 
complex phase of all the spurions $F^A$ is set to zero.
These are the two main situations where we still keep CP symmetry,
where genericaly, because of $u=<{\rm Tr}\phi^2> \not= 0$, the vacuum 
will break CP spontaneously. We will find such vacua in sections 2 and 3.
But for complex bare masses and supersymmetry breaking parameters 
with relative complex phases among them, we do not have CP invariance in the 
bare Lagrangian.
In those cases, we will get effective Lagrangians 
at low energy which are not CP invariant.

\subsection{The Low Energy Effective Couplings}

In (I), a general procedure to compute the Seiberg-Witten periods 
was introduced, 
based on the uniformization of the elliptic curve associated to the theory
\footnote{
This method has also been considered independently in 
\cite{ferrari, bfmasas}.}. 
In this way we have a map from the ${\bf C}/ \Lambda$ lattice to the 
$(x,y)$ variables of the Seiberg-Witten curve
 through the $\wp (z)$ Weierstrass elliptic function. The 
variables are given by
\be
x= 4 \wp (z) + {u \over 3} + {\delta_{N_f, 3} \over 192} \Lambda_3 ^2, 
\,\,\,\,\,\,\,\ y=4 \wp ' (z). 
\label{variables}
\ee
The Seiberg-Witten abelian differential can be written as 
\be
\lambda_{SW} = {{\sqrt 2} \over 8\pi}{dx \over y} (2u - (4-N_f)x) 
+ {dx \over y} \sum_{n=1} ^{N_p} {r_n \over x-x_n}.
\label{lambda}
\ee
In this equation, $N_p$ denotes the number of poles (which depends on $N_f$), 
located at $x_n$, and the coefficients $r_n$ are given by
\be
r_n= 4 \wp'(z_n) [{\rm Res}_{x=x_n} \lambda_{SW}],
\label{coef}
\ee
where the $z_n$ correspond to the poles $x_n$ through (\ref{variables}). The 
residues can be written as linear combinations of the masses, and 
this defines the coefficients $S_n^f$ as follows:
\be
{\rm Res}_{x=x_n} \lambda_{SW}= {1 \over 2 \pi i} \sum_{f=1}^{N_f} S_{n}^f 
{m_f \over {\sqrt 2}}.
\label{residuos}
\ee
The explicit values of $\wp'(z_n)$ and $S_n^f$ can be found in (I). 

Denoting 
$a_1 = a_D$, $a_2= a$, one can integrate (\ref{lambda}) over the basic 
homology cycles to obtain the explicit expressions
\bea
a_i &=& {{\sqrt 2} \over \pi} \Big( (4-N_f)\zeta({\omega_i \over 2}) +
 ({ N_f+2 \over 24})u \, \omega_i - ({ \d_{N_f, 3} 
\Lambda_3^2 \over 1536}) \omega_i \Big) \nonumber\\
&+& 2i 
\sum_{n=1} ^{N_p} [{\rm Res}_{x=x_n} \lambda_{SW}] \big[\omega_i \zeta (z_n)-
2 z_n \zeta \big({\omega_i \over 2} \big)  \big],
\label{as}
\eea
where $\zeta(z)$ is the Weierstrass zeta-function, and $\omega_i$ are 
the periods of the abelian differential $dx/y$. Expressions for
 the derivatives of $\lambda_{SW}$ w.r.t. the masses 
can be easily found, and these 
give in turn
\be
\Big({\partial a_i \over \partial m_f}\Big)_u
={1 \over {\sqrt 2}}\sum_{n=1} ^{N_p} {S_n^f \over \pi i} 
\big[\omega_i \zeta (z_n)-
2 z_n \zeta \big({\omega_i \over 2} \big)  \big].
\label{parcialmasa}
\ee
Using the Riemann bilinear relations, 
one can obtain an equation for the dual mass (I),
\be
m_D^f = {\sqrt 2}\Big( {\partial {\cal F} (a, m_f) \over \partial m_f} \Big)_a
= \sum_{n=1}^{N_p} S^f_n \int_{x^-_n}^{x^+_n} \lambda_{SW},
\label{mD}
\ee
where ${\cal F} (a, m_f)$ is the Seiberg-Witten prepotential for the massive
theories, and the points $x^+_n$ and $x^-_n$ ($n=1,\cdots,N_p$) are the simple
 poles of $\lambda_{SW}$
at each of the two Riemann sheets. The ${\sqrt 2}$ factor arises 
because it is more convenient to use $m_f/{\sqrt 2}$ as mass variables, 
given our normalizations. The expression (\ref{mD}) is defined 
up to an $a$-independent constant, that we will set to zero. 
As it has been remarked in \cite{dhoker}, 
this expression has to be 
regularized in order to get a finite result, but in our approach via elliptic 
funtions this can be easily implemented (see (I, 3.30)). We make the simple 
choice $z_n^{-}=-z_n^{+}$, and denote $z_n^{+}=z_n$. We then obtain:
\bea
m_D^f &=& \sum_{n=1} ^{N_p} S_n^f \Bigg( {{\sqrt 2} \over \pi} 
\Big( (4-N_f)\zeta(z_n) +
 ({ N_f+2 \over 12})u \, z_n  - ({ \d_{N_f, 3} 
\Lambda_3^2 \over 768})z_n  \Big) \nonumber\\
&+& [{\rm Res}_{x=x_n} \lambda_{SW}] \big[4 z_n \zeta (z_n)-
2 {\rm log} \sigma(2z_n)  \big] \Bigg) \nonumber\\
&+& \sum_{n \not= m}^{N_p} S_n^f [{\rm Res}_{x=x_m} \lambda_{SW}] 
\big[4 z_n \zeta (z_m)-
2 {\rm log} {\sigma(z_n-z_m) \over \sigma(z_n+z_m)} \big],
\label{masdex}
\eea
where $\sigma(z)$ is the Weierstrass sigma-function. 
 
Using these expressions it is easy to check that 
\be
a_D {\partial a \over \partial u}-a {\partial a_D \over \partial u}-
\sum_{f=1}^{N_f}{m_f \over {\sqrt 2}} {\partial m_D^f \over \partial u}= 
-i{4-N_f \over 4 \pi},
\label{matone}
\ee
which is essentially the relation derived in \cite{amz, matones, dhoker}.

Once supersymmetry is softly broken with dilaton and mass spurions as in (I), 
the low-energy 
effective potential is described in terms of a series of couplings. These 
can also be written in terms of elliptic functions using (I, 3.36-3.40). We 
will choose the ``electric" description, although the magnetic one can 
be obtained either by direct computation or by using the generalized 
duality transformations in (I, 3.14):
\bea
\tau^{aa}&=&\frac{\partial^2 {\cal F}}{\partial a^2}=
 {{\omega_1}\over {\omega_2}}, \nonumber\\
\tau^{af}&=& \frac{\partial^2 {\cal F}}{\partial a \partial m_f}=
{2 \over \omega_2} \sum_{n=1} ^{N_p} S_n^f z_n, \nonumber\\
\tau^{0a}&=& \frac{\partial^2 {\cal F}}{\partial a \partial s}=
{{\sqrt 2}\over \omega_2}(4-N_f) , \nonumber\\
\tau^{0f} &=& \frac{\partial^2 {\cal F}}{\pa m_f \pa s}=-{\sqrt 2}(4-N_f) 
\sum_{n=1} ^{N_p} {S_n^f \over \pi i}\big[\zeta (z_n)-
 {2 z_n \over \omega_2} \zeta \big({\omega_2 \over 2} \big)  \big], \nonumber\\
\tau^{00}&=& \frac{\partial^2 {\cal F}}{\pa s^2}= i {(4-N_f)^2 \over \pi} 
\Big({u \over 12}- {2 \over \omega_2} \zeta \big({\omega_2 \over 2}\big) \Big) 
+ i \delta_{N_f,3} {\Lambda_3^2 \over 256 \pi}, \nonumber\\
\tau^{fg}&=& \frac{\partial^2 {\cal F}}{\pa m_f \pa m_g}=-\sum_{n=1} ^{N_p} 
{S_n^f S_n^g \over \pi i} \big[ {\rm log} \sigma (2 z_n) - {4 z_n^2 \over 
\omega_2} \zeta \big({\omega_2 \over 2}\big) \big]\nonumber \\
&+&\sum_{n \not= m}^{N_p}{S_n^f S_n^g \over \pi i} \big[ {\rm log} 
{\sigma( z_n-z_m)\over \sigma(z_n+ z_m)}+  {4 z_n z_m \over 
  \omega_2} \zeta \big({\omega_2 \over 2}\big) \big].
\label{acoplos}
\eea

In $\tau^{fg}$, the divergences have been subtracted following the method
explained in (I).
Notice that the $z_n$ variables are defined up to a shift of an integer linear 
combination of the periods ${\omega}_i$, $i=1,2$, and the $SL(2,{\bf Z})$ 
group acts in a natural way on the ${\omega}_i$, $i=1,2$. One can easily 
check from the above expressions for the $a_i$ that these two 
sets of transformations 
(shifts in $z_n$ and $SL(2,{\bf Z})$ transformations) combine together 
to give the inhomogeneous duality transformations of the Seiberg-Witten 
model for the massive theories \cite{swtwo}. In fact,
 the generalized duality 
transformations (I, 3.14) for the couplings can be derived 
in terms of these 
geometrical transformations by using the explicit expressions 
in (\ref{acoplos}).

\subsection{The Squark Condensates}

The squark condensates can be exactly computed in terms of these couplings, 
starting from the expression for the bare Lagrangian (\ref{lagrangiano})
and regarding the supersymmetry breaking parameters $F^f$ as source terms. 
We have:
\be
\langle {\tilde q}_f q_f \rangle = -{1 \over {\sqrt 2}} 
{\partial V_{{\rm eff}} \over 
\partial F_f} = {1 \over {\sqrt 2}} \Bigl(b_{fB}-
{b_{af} b_{aB} \over b_{aa}}\Bigr){\overline F}_B +
\Big( S_i^f -{ b_{af} \over b_{aa}} \Big) < h_i {\tilde h}_i> , 
\label{squark}
\ee 
where we follow the notation in (I). Namely, $b_{\alpha \beta} = 
{\rm Im} \tau_{\alpha \beta}/4 \pi$, where $\alpha$, $\beta=a,0,f$, 
$A$,$B=0,f$, and $h_i$, ${\tilde h}_i$ are the scalar components of the 
$i=1, \cdots, k$ hypermultiplets becoming massless near a singularity, with 
baryon numbers $S_i^f$. Using this expression we can compute the squark 
condensate at the minimum of the effective potential, and in this way 
we have an additional indication of the ``Higgsing" effect of the 
soft breaking terms.

\section{$N_f=1$ Vacuum Structure}
\setcounter{equation}{0}

 In this section we study the phase diagram and vacuum structure of the 
softly broken $N=2$ QCD with one massive hypermultiplet. The massless 
case was studied in \cite{softdos}, for the soft breaking induced solely by 
the dilaton spurion. It was found that the theory has generically two 
degenerate minima with dyon condensation, and at a certain value 
of the supersymmetry breaking parameter there is a first order 
phase transition
to a single minimum. This new vacuum 
shows simultaneous condensation of mutually non-local states and has 
a natural interpretation in terms of oblique confinement \cite{ob}. In fact, 
the two dyons condensing there have opposite electric charges and magnetic 
number $n_m=1$. We will see more examples of these ``bound states" 
in the massive models. 

We will study in some detail the theories with arbitrary hypermultiplet 
mass $m$ and one of the two spurion fields turned on. Two different cases
 will be considered, with parameter space given by $(m, f_0)$ and $(m, f_m)$, 
respectively. 
For simplicity, the supersymmetry breaking parameters and the mass will 
be real and positive, although the most interesting physical phenomena are 
already 
captured in these simple cases. In the first subsection we will give the 
explicit expressions for the monopole condensate and the effective potential. 
In the second subsection we will analyze the theory with the dilaton spurion 
breaking parameter $f_0$. In the third subsection we will focus on the 
special case $m=0$ and with only the mass spurion breaking 
the supersymmetry down to $N=0$. Finally, in the last subsection we will 
study the phase structure of the theory for general $m$ and $f_m$.

\subsection{Condensates and effective potential}

The general expression for the effective potential of the softly broken theory 
has been given in (I, 4.7). For $N_f=1$ there is only one massless BPS state
near each singularity \cite{swtwo}, and one has:
\bea
V_{{\rm eff}}&=&\Bigl({b_{aA} b_{aB} \over b_{aa}}-b_{AB}\Bigr) 
\Bigl({1 \over 2}D_A D_B + F_A {\overline F}_B \Bigr) + 
{b_{aA} \over b_{aa}}D_A  (|h|^2-|{\widetilde h}|^2) 
\nonumber \\ 
&+& { \sqrt{2} b_{aA}\over b_{aa}} \Bigl( F_A
  h{\widetilde h} + {\overline F}_A 
{\overline h}{\overline {\widetilde h}} \Bigr) + 
{1 \over 2  b_{aa}} (|h|^2 + |{\widetilde h}|^2)^2
\nonumber\\
&-& D_m S (|h|^2-|{\widetilde h}|^2) + 2|a + 
S {m \over {\sqrt 2}}|^2(|h|^2+|{\widetilde h}|^2) 
\nonumber 
\\
&-&
\sqrt{2} \Bigl( S F_m h {\widetilde h} + 
S{\overline F}_m {\overline h}{\overline {\widetilde h}} \Bigr).
\label{twovi}
\eea
The $A,B=0,m$ labels refer to the mass and dilaton spurions.  
We will set $D_0=D_m=0$ for simplicity. Anyway, because of the $SU(2)_R$ 
covariance of (\ref{twovi}), the monopole solution which 
minimizes the potential for a general spurion configuration is just the 
adequate $SU(2)_R$ rotation of the solution 
we will obtain here. 
 
Minimizing w.r.t the monopole fields one obtains the equations
\be
{\partial V_{{\rm eff}} \over \partial {\overline h}}={1 \over b_{aa}}
\big( |h|^2+|{\widetilde h}|^2 \big)h+2|a + S {m\over {\sqrt 2}}|^2 h +
 {{\sqrt 2} \over
b_{aa}}\Big(b_{aA}{\overline F}_A-
 S b_{aa}{\overline F}_m \Big){\overline {\widetilde h}}=0,
\label{fivei}
\ee
\be
{\partial V_{{\rm eff}} \over \partial {\overline {\widetilde h}}}={1 \over
b_{aa}}\big( |h|^2+|{\widetilde h}|^2 \big){\widetilde h} + 
2|a+ S {m \over {\sqrt 2}}|^2 {\widetilde h
}+ {{\sqrt 2} \over b_{aa}}\Big(b_{aA}{\overline F}_A -
S b_{aa}{\overline F}_m\Big) {\overline h}=0.
\label{fiveii}
\ee
Multiplying (\ref{fivei}) by ${\overline h}$, (\ref{fiveii}) by 
${\overline {\widetilde h}}$ and
subtracting we obtain:
\be
\Big( {1 \over b_{aa}}( |h|^2+|{\widetilde h}|^2 ) 
 + 2|a+ S {m \over {\sqrt 2}}|^2 \Big) ( |h|^2-|{\widetilde h}|^2)=0.    
\label{iii}
\ee
If $|h|^2+|{\widetilde h}|^2 >0$, then it follows 
from (\ref{iii}) that $|h|^2=|{\widetilde h}|^2$. We can fix the 
gauge and absorb 
the phase of $F_{0}$ such that 
\be
h=\rho , \,\,\,\,\ {\widetilde h}= \rho  {\rm
e}^{i\beta}, \,\,\,\,\ F_{0}=f_{0}
\label{fiveiv}
\ee
without loss of generality. Here, $\rho$ and $f_0$ are real and positive.
 Substituting (\ref{fiveiv}) in
(\ref{fivei}) leads to:
\be
\rho^2+ b_{aa}|a+ S {m  \over {\sqrt 2}}|^2+ {1 \over {\sqrt
2}}\Big(  b_{a0}f_0 + {\overline F}_m (b_{am} -S b_{aa}) \Big) 
{\rm e}^{-i \beta} =0,
\label{fiveviii}
\ee
 apart form the trivial solution $\rho=0$. This implies that 
\be
{\rm e}^{-i \beta}=-{ b_{a0}f_0 + {\overline F}_m (b_{am} -S b_{aa}) \over 
|b_{a0}f_0 + {\overline F}_m (b_{am} -S b_{aa})|}, 
\label{fase}
\ee
and the monopole condensate is given by
\be
\rho^2=-b_{aa}|a+ S {m  \over {\sqrt 2}}|^2 + 
{1 \over {\sqrt
2}}|b_{a0}f_0 + {\overline F}_m (b_{am} -S b_{aa})|. 
\label{monopolo}
\ee
Depending on the phase of $F_m$ we will have different situations. If one 
writes 
$F_m =f_m {\rm e}^{i \delta}$, it is easy to see that the most 
favourable situation 
to have a condensate corresponds to 
\be
{\rm cos}\delta^{*} = {\rm sign} \Big( b_{a0}(b_{am}-S b_{aa}) \Big),
\label{coseno}
\ee
and the opposite is the less favourable. When the solution 
(\ref{monopolo}) is 
introduced in (\ref{twovi}) we obtain the final expression 
of the
effective potential as a function on the $u$-plane:
\be
V_{{\rm eff}}= -{ 2\over b_{aa}} \rho^4 
-\Bigl({b_{aA} b_{aB} \over b_{aa}}-b_{AB}\Bigr) 
F_A {\overline F}_B. 
\label{finalef}
\ee
The second term in (\ref{finalef}) is the cosmological constant, and as we 
already know from (I) it is a monodromy invariant. Near each singularity 
one must include the corresponding condensate, even when 
the condensates overlap. The approximation breaks 
down when one of 
the condensates attains the singularity associated to other massless state 
\cite{softdos}. 
This will typically happen when the value of the supersymmetry breaking 
parameter is comparable to the value of the dynamically generated scale of 
the theory, and one should include then higher order derivative corrections. 
As we will see, in the massive theories the range of validity crucially 
depends on the value of the bare mass, since the  singularities on the 
$u$-plane move 
when $m$ is increased.       

Recall that one can shift the $a$ variables by integer 
multiples of the residues. The numerical analysis of the expressions for 
the $a$, $a_D$ variables in the $N_f=1$ case tells us that one can 
define the variables around each singularity in such a way that 
the corresponding massless state has $S=0$. This implies that $S$ 
must be an integer (as $S^n_f=-1$ in this case), and agrees with the 
general considerations in \cite{ferraprl} and the results of \cite{andreas}. 
This redefinition of $a$ also sets $S=0$ in the term $b_{am} -S b_{aa}$, as 
one can easily check using the duality transformations in (I).

\subsection{Turning on $f_0$: CP Symmetry and the 
Argyres-Douglas Point}

In the analysis of the theory we will normalize the dynamical scale 
as $\Lambda_1^6 = 256/27$. Before proceeding to the study of 
the phase structure of the model, it is 
important to know the evolution of the singularities on the $u$-plane as the 
mass is turned on. 
\begin{figure}
\epsfxsize=6cm
\centerline{\epsfbox{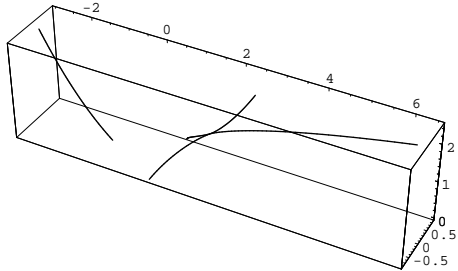} }
\caption[]{Evolution of the singularities on the $u$-plane for $0<m<2.5$. }
\figlabel\singu
\end{figure}
This is shown in \fig\singu\ for a real mass 
$0<m<2.5$, represented on the vertical 
axis. At $m=0$ we start from the theory with one massless hypermultiplet, 
where the singularities on the $u$-plane are related by the non-anomalous 
${\bf Z}_3$ symmetry. They correspond to BPS states with quantum 
numbers $(n_e, n_m)=(0,1)$, $(1,1)$ and $(2,1)$, and with the 
above normalization for the dynamical scale these singularities are located 
respectively at $u_1={\rm e}^{-i\pi/3}$, $u_2={\rm e}^{i\pi/3}$ 
and $u_3=-1$.  
The mass terms explicitly break the ${\bf Z}_3$ symmetry and we have the 
following evolution: for a real mass, the $(1,0)$ and $(1,1)$ states approach 
each other and collapse on the real $u$-axis when $m =m_c= 
3 \Lambda_1/4 \sim 
1.09$.
This 
is the Argyres-Douglas point \cite{ad} discovered in \cite{apsw}, and as 
we will see it 
plays an important role in the phase structure of the model. At this point 
the two collapsing states are simultaneously massless and the theory 
describing this situation is an $N=2$ superconformal theory. For $m>m_c$, one 
of the two collapsing singularities corresponds now to a massless  
elementary quark and 
the other to a $(0,1)$ monopole. The change of quantum numbers is due to the 
conjugation of monodromies after the singularities collide 
\cite{apsw, bfmasas}. 
When $m$ is increased the quark decouples and 
the remaining singularities locate on approximate symmetric positions with 
respect to the imaginary $u$-axis. The theory becomes pure $N=2$ Yang-Mills 
with the singularity structure unveiled in \cite{swone}.  

We will then study the vacuum structure of the theory as we change two 
parameters: the bare quark mass $m$ and the dilaton spurion $f_0$. It 
is important 
to notice that both of them explicitly break the ${\bf Z}_3$ 
symmetry of the massless theory, and this will be apparent in our 
solution. 

For small values of the mass, once the 
supersymmetry breaking parameter is 
turned on, the vacuum structure is very similar to the one found in the 
massless $N_f=1$ theory. For real values of the bare mass, the effective 
potential (\ref{finalef}) (once all the condensates are included) has a 
CP symmetry, which 
relates $u \rightarrow {\bar u}$. There are 
generically two degenerate minima on the $u$-plane associated to the 
condensation 
of the $(1,0)$ monopole 
and $(1,1)$ dyon, and the CP symmetry is spontaneously 
broken. The remaining dyon with quantum numbers $(2,1)$ develops a very 
tiny condensate and does not produce a minimum (in fact, the cosmological 
constant is smooth at the $(2,1)$ singularity). The minima move away from 
the singularity as the 
supersymmetry breaking parameter is increased, and the effective theta angle 
gives opposite electric charges to the condensates. Therefore the two 
condensing states are in fact dyons with opposite electric 
charges at conjugate points on the $u$-plane. 
The situation is exactly like the one for $m=0$, and there is a smooth 
connection 
between these two situations. We also find a first order phase transition for 
a critical value of $f_0$, with the structure described in \cite{softdos}. The 
CP symmetry on the $u$-plane is restored and we have a simultaneous 
condensate of two mutually non-local states which could be interpreted as 
a bound state with zero electric charge and magnetic charge $n_m=2$. This 
suggests an interpretation of this vacuum in terms of oblique confinement. 
When $f_0$ is still increased we reach a maximum value and our approximation 
breaks down. It is interesting to notice that in these two phases, and 
according to (\ref{squark}), we have 
a non-zero squark condensate. The couplings between the Higgs field and the 
squarks appearing in the soft breaking terms of (\ref{lagrangiano}) favour 
a non zero VEV for ${\tilde q}q$.   

  \figalign{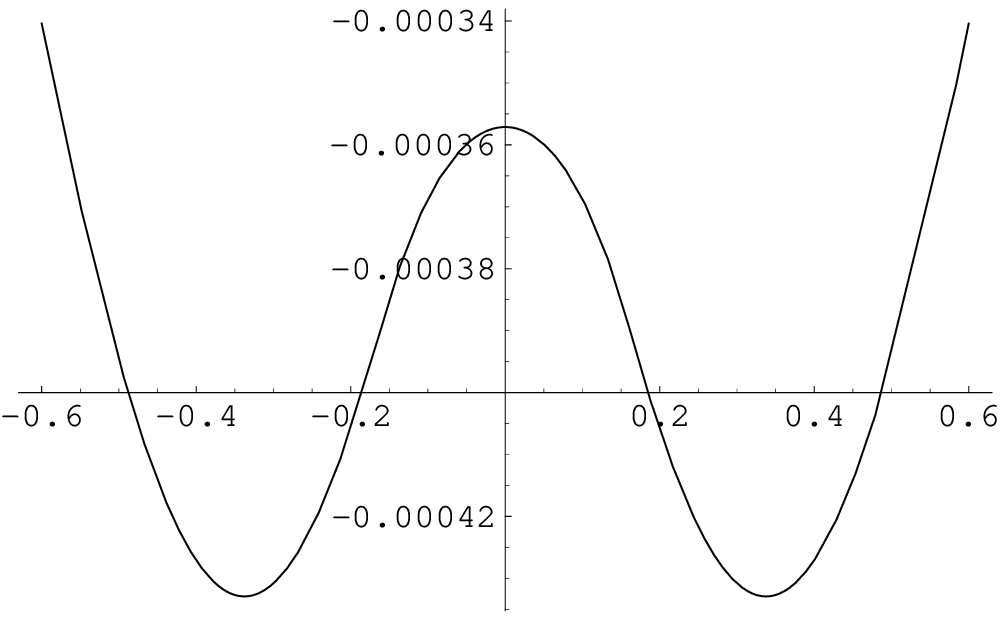}{\founo}{The two minima of the effective potential 
for $m=0.5$ and $f_0=0.15$ along $u=0.82 + i y$. 
}{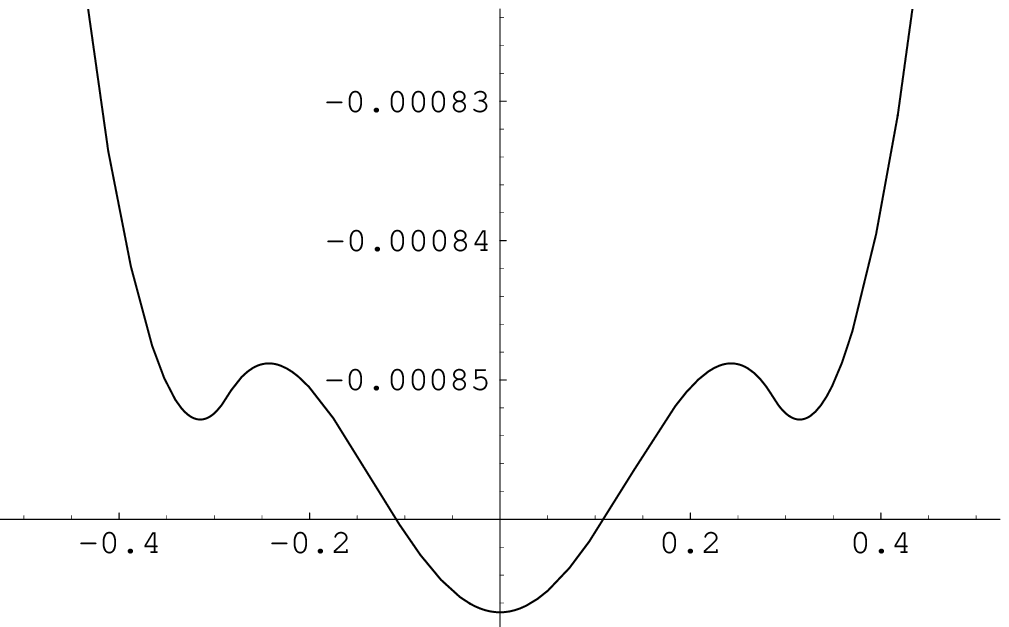}{\fodos}{The minimum of the effective potential on the real 
$u$-axis for $m=0.5$ and $f_0=0.21$, along $u=0.76 + i y$. }

A typical situation is shown in \fig\founo , where the value of the 
supersymmetry breaking parameter is below the transition point ($f_0=0.15$). 
When $f_0$ reaches a critical value around $f_0 \sim 0.2$, the new absolute 
minimum takes place on the real $u$-axis and the CP symmetry 
$u \rightarrow {\bar u}$ is restored. This is shown in \fig\fodos. 

As the mass increases, the monopole and $(1,1)$ dyon singularity approach 
each other and the maximum allowed value for $f_0$ decreases. When the 
singularities are very close to each other, for small values of the 
supersymmetry breaking parameter a singularity develops in the effective 
potential and our approximation breaks down. For $m=0.5$, for instance, 
the maximum allowed value is $f_0 \sim 0.21$, to be compared with the 
maximum value $f_0 \sim 0.8$ for the massless case \cite{softdos}. We then see 
that the range of validity of our approximation becomes smaller as we approach 
the Argyres-Douglas point.
At the same time, the ``window" in which the theory has a single minimum 
narrows and finally dissapears at a certain value of the mass $m \sim 0.8$. 
The theory has from this moment on a single phase with two degenerate minima 
(for the allowed range of values for $f_0$).  

At $m=m_c$ we have a singular situation: the description in terms of 
the effective potential breaks down for {\it any} value of $f_0$. The monopole 
and dyon condensates have a cusp singularity and our description 
of the physics in terms of an effective potential is no longer valid. 
This is natural if we take into account that at this point we have a conformal 
field theory with no mass scale, and all the higher order corrections in 
powers of $f_0$ become equally important.

For $m>m_c$ the vacuum structure of the theory completely changes. After the 
collapsing of the singularities one of them is naturally interpreted 
as an elementary quark becoming massless (the adequate 
variable is then the electric one, $a$), and the other as a $(0,1)$ monopole. 
When we break supersymmetry down to $N=0$ with the dilaton spurion, there is 
a quark condensate around the singularity corresponding to the decoupling 
state. We also have a monopole condensate around the monopole singularity, 
while the $(2,1)$ condensate is still very tiny.     
\begin{figure}
\centerline{
\hbox{\epsfxsize=6cm\epsfbox{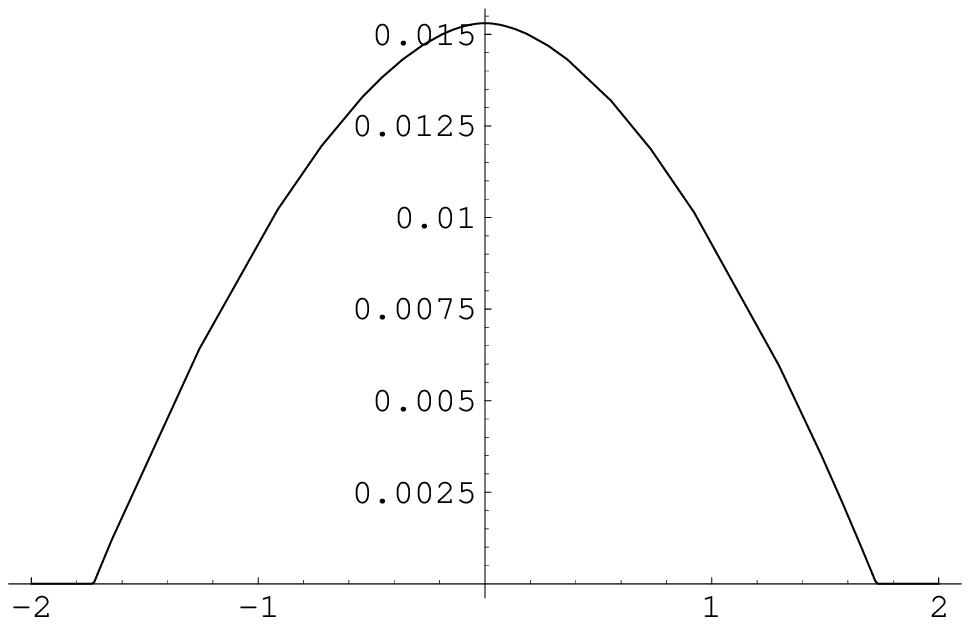}}\qquad
\hbox{\epsfxsize=6cm\epsfbox{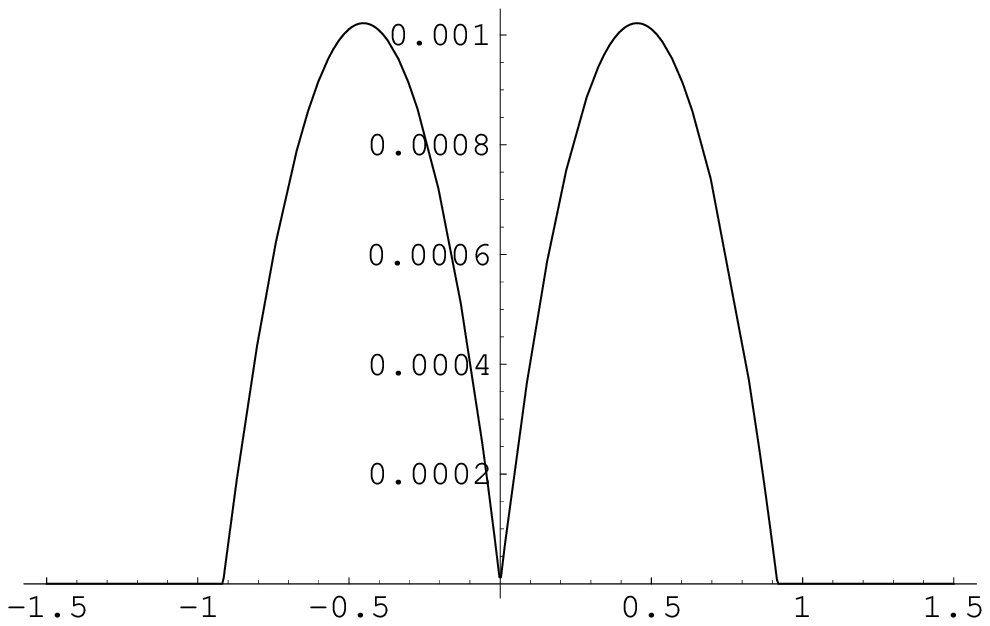}} }
\caption[]{Monopole condensate (left) and quark condensate (right) for 
$m=2$, $f_0=0.4$. Both are centered around the corresponding singularity 
and plotted along the imaginary $u$ direction.}
\figlabel\quarkcon
\end{figure}
The monopole and quark condensates are plotted in \fig\quarkcon\ for $m=2$ 
and $f_0=0.4$. For these values of the mass the monopole and quark 
singularity are located at $u \sim 2.3$ and $u \sim 4.2$, respectively. 
The figures  
are centered around the corresponding singularities and plotted along the 
imaginary 
direction. We see that the quark condensate is one order of magnitude 
smaller than the monopole one. In fact, the minimum of the effective potential 
is located on the real $u$-axis and close to the monopole singularity, 
exactly like in \cite{soft}. As 
the effective theta angle is zero in this region, the minimum corresponds 
to a pure monopole condensate and we have a confining phase strictly speaking. 
In fact, the squark condensate (\ref{squark}) vanishes at this phase. 
\begin{figure}
\epsfxsize=6cm
\centerline{\epsfbox{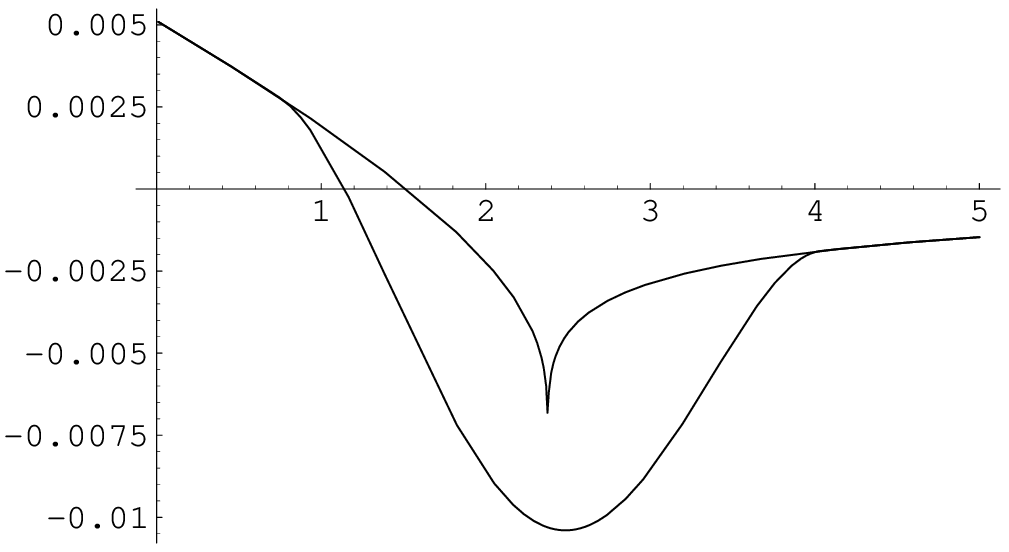} }
\caption[]{Cosmological constant (top) and effective potential (bottom) for 
$m=2$ and $f_0=0.4$ along the real $u$-axis. }
\figlabel\fovacquark
\end{figure} 
The effective potential as well as the cosmological 
constant are shown in \fig\fovacquark, for the same values of 
the parameters, with the mentioned above behaviour. Notice that the 
cosmological constant has no 
cusp at the quark singularity, and this is already a sign that this state 
won't dominate the vacuum structure. This is also the case in the models 
studied in \cite{soft, softdos}. As the mass increases the quark condensate 
becomes smaller and the monopole condensate bigger, for the same 
value of the supersymmetry breaking parameter. In the decoupling limit of very 
large masses we approach the softly broken $SU(2)$ Yang-Mills theory studied in 
\cite{soft}. The softly broken $N_f=1$ theory with $m>m_c$ is then smoothly 
connected to the 
pure Yang-Mills case, as expected. 
\begin{figure}
\epsfxsize=5cm
\centerline{\epsfbox{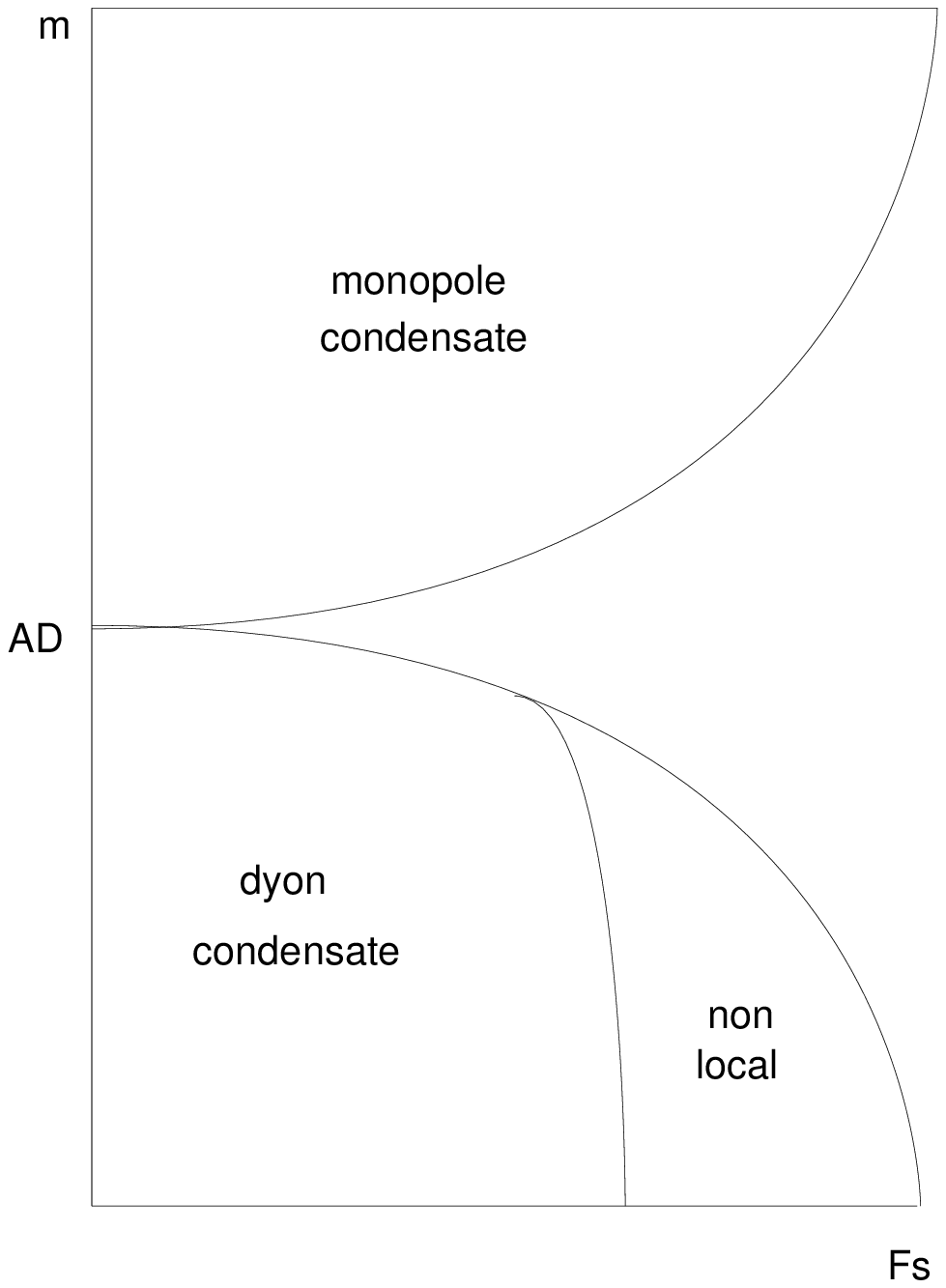} }
\caption[]{Phase structure of the $N_f=1$ theory for $m$, $f_0$ (denoted as 
Fs). AD denotes the critical value of the mass $m_c$, where the 
Argyres-Douglas point occurs. }
\figlabel\diagrama
\end{figure}

The phase structure of this theory can be summarized in \fig\diagrama. 
We see that there are two different behaviours, depending on wether 
$m<m_c$ or $m>m_c$ (this critical value of $m$ is denoted 
AD in the diagram). For each value of $m$ there is a maximum allowed value 
of $f_0$ which goes to zero as $|m-m_c|$. When $m<m_c$ there are two 
different phases for small values of $m$, separated by 
a first order phase transition (a sudden jump of the position of the minimum 
on the $u$-plane). One of them is characterized by dyon 
condensation, and the other by the simultaneous condensation of mutually 
non-local states, and corresponding possibly to a oblique confinement phase 
(this phase is denoted as ``non local" in the diagram). 
As $m$ approaches $m_c$ from below, the second phase dissapears and we are 
only left with the first one. When $m>m_c$, the theory is in a different 
phase, characterized by monopole condensation and confinement, and is smoothly 
connected to the softly broken Yang-Mills theory.     
     
\subsection{Turning on $f_m$ at $m=0$: ${\bf Z}_3$ Symmetry
 and a New Phase}     

In this subsection we will study the vacuum structure of the theory in which 
$m=0$ but we turn on a supersymmetry breaking parameter coming 
from the mass spurion superfield, $f_m$. Notice that 
the couplings of the spurion are encoded in the 
holomorphic dependence on the mass, regarded as a $U(1)$ $N=2$ 
vector multiplet, but we can set the scalar component of this hypermultiplet 
to zero. The breaking terms in the 
microscopic Lagrangian only involve the squarks, and as $m=0$  
the ${\bf Z}_3$ discrete symmetry of the $N=2$ supersymmetric theory 
is preserved (for the squarks have zero $R$ charge). We also 
have the CP symmetry $u \rightarrow {\bar u}$. In fact, 
as we will see in a moment, these global discrete symmetries govern 
at a large extent the 
dynamics of the theory. The ${\bf Z}_3$ symmetry is also 
preserved  when 
we softly break $N=2$ down to $N=1$ supersymmetry with a mass term for the 
$N=1$ chiral superfield $\Phi$ in the $N=2$ vector multiplet. In this case, 
the minima are locked at the three singularities \cite{swtwo}.    

\begin{figure}
\centerline{
\hbox{\epsfxsize=6cm\epsfbox{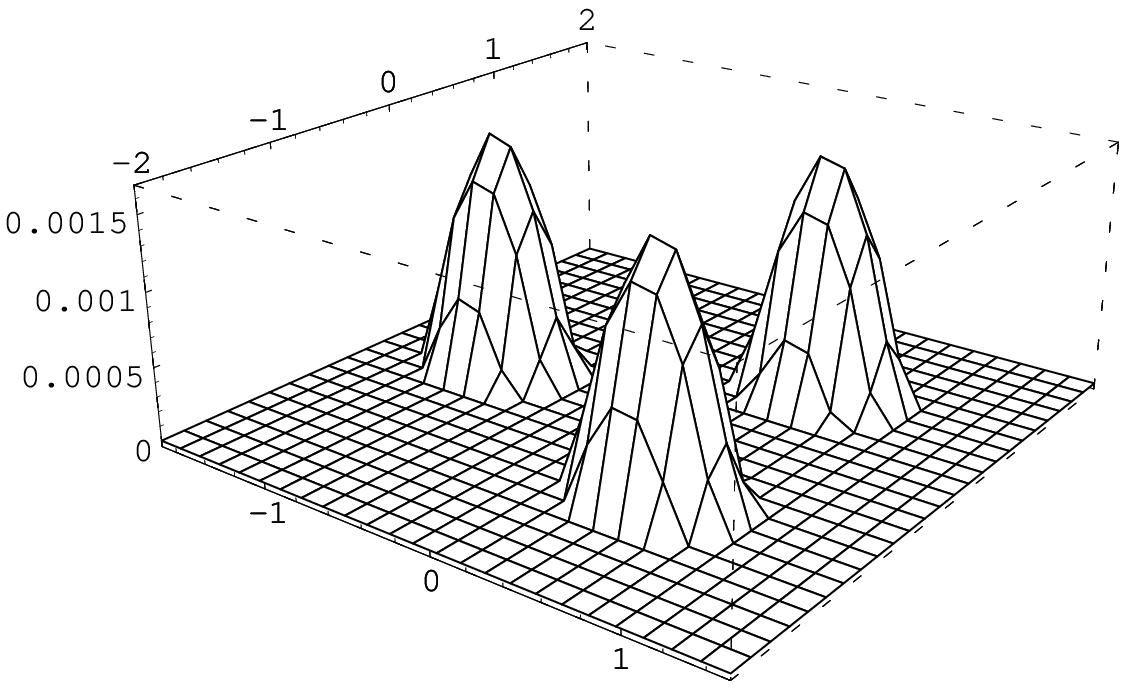}}\qquad
\hbox{\epsfxsize=6cm\epsfbox{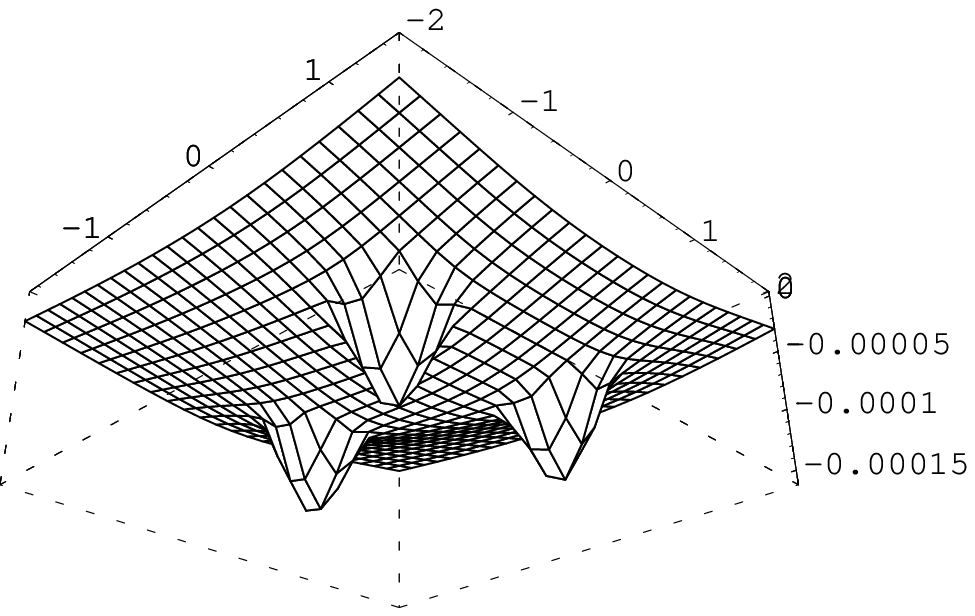}} }
\caption[]{Condensates (left) and effective potential (right) on the 
$u$-plane for 
$m=0$, $f_m=0.14$.}
\figlabel\plane
\end{figure}

For small values of the supersymmetry breaking parameter, the three 
BPS states condense around the corresponding singularities. This gives three 
degenerate minima spontaneously breaking the ${\bf Z}_3$ 
symmetry. They are located on the segments going from the three singularities 
to the origin of the $u$-plane, as required by the symmetry of the model. 
In fact they move away towards $u=0$, and for $f_m>0$ one has $|u|<1$ for 
the three minima. This generic situation is shown in \fig\plane. It is 
important to notice that all the three states have an electric charge given by 
the effective theta angle through Witten's effect. For the $(0,1)$ and $(1,1)$ 
states we have the same situation than in the dilaton spurion case: 
they have opposite electric charges. The $(2,1)$ state acquires an electric 
charge as the minimum moves towards the origin. We also have a nonzero 
squark condensate (\ref{squark}) at the minima which increases with $f_m$, 
as expected. The 
theory is then ``Higgsed" as we turn on the mass spurion supersymmetry breaking 
parameter.   

As in the previous cases, there is a maximum allowed value for $f_m \sim 1.4$, 
but before we reach this value a phase transition occurs: as the three minima 
approach each other they begin to overlap (as in \cite{softdos}) and a new 
minimum develops at $u=0$. For $f_m \sim 0.93$ the minimum suddenly jumps to 
the origin and the ${\bf Z}_3$ symmetry is restored. There is 
there a simultaneous condensate of {\it three} mutually non-local 
states, which can be interpreted 
perhaps as a bound state with nonzero electric charge (in contrast to the 
bound state associated to two mutually non-local states). The electric charge 
of this object is associated to the $(2,1)$ state and would be $1/2$, while   
its magnetic charge would be $n_m=3$. This is a new different phase appearing 
in the softly broken $N=2$ QCD models. 
           
\begin{figure}
\centerline{
\hbox{\epsfxsize=6cm\epsfbox{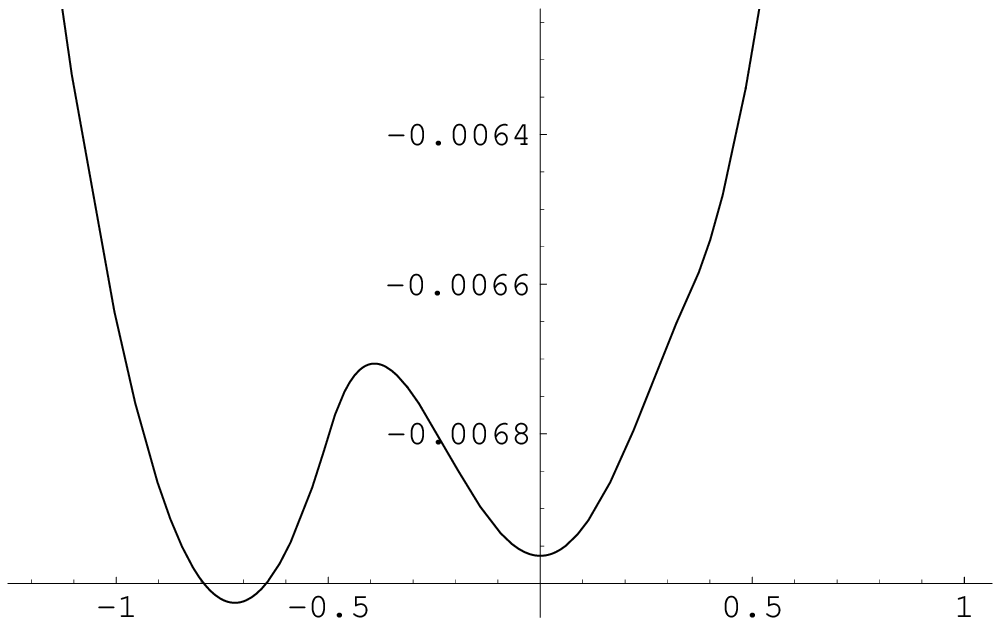}}\qquad
\hbox{\epsfxsize=6cm\epsfbox{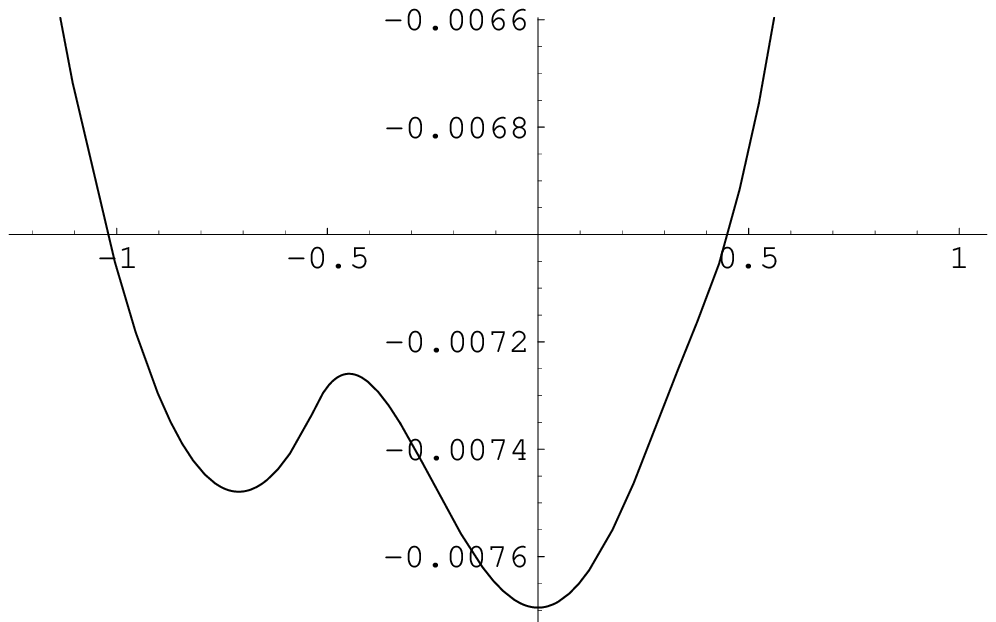}} }
\caption[]{Effective potential along the real $u$-axis for $m=0$ and 
$f_m=0.92$ (left) and $f_m=0.95$ (right).}
\figlabel\trans
\end{figure}

\begin{figure}
\epsfxsize=6cm
\centerline{\epsfbox{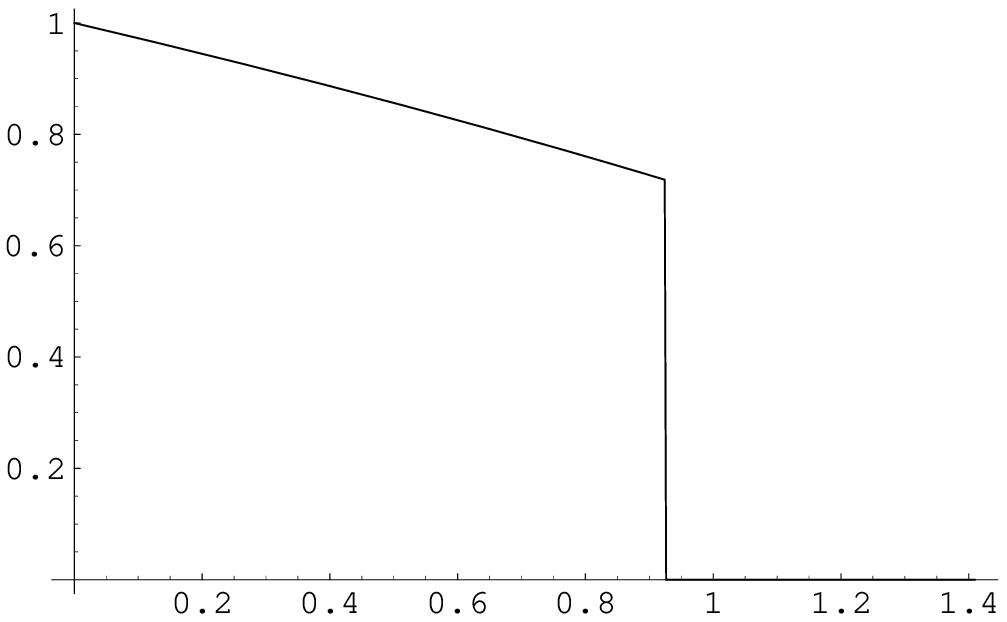} }
\caption[]{The value of $|u|$ (vertical axis) at the minima as a function of 
$f_m$ (horizontal axis). The phase 
transition to $u=0$ takes place for $f_m \sim 0.93$.}
\figlabel\min
\end{figure}

This phase transition is clearly shown in \fig\trans, where the effective 
potential is plotted along the real $u$-axis. One of the three degenerate 
minima (the one associated to the $(2,1)$ state) appears on the negative real 
axis. As $f_m$ passes through the critical value, the phase transition 
takes place and we are in the new phase with a minimum at $u=0$ and 
three simultaneous condensates. This is the phase we find until $f_m$ reaches 
the maximum allowed value. It is also illustrative to consider the 
evolution of $|u|$ at the minima as we turn on the supersymmetry breaking 
parameter $f_m$. This is plotted in \fig\min, where the sudden change in the 
position of the minimum is apparent.

\subsection{The Phase Structure for $m$, $f_m$}

The above picture changes when a real mass $m$ is introduced. As this 
explicitly breaks the ${\bf Z}_3$ symmetry, the only constraint on the 
dynamics of the theory comes from the CP symmetry. The mass term 
makes the $(2,1)$ minimum less favourable than the other two, and for 
small values of $m$, $f_m$ we have 
then two minima (associated to the $(0,1)$ and $(1,1)$ states) 
spontaneously breaking the CP symmetry. The situation 
is similar to the one found for the dilaton spurion breaking. As $f_m$ is 
increased, we have an overlapping of the condensates and a new minimum appears 
on the real $u$-axis with a first order phase transition, again similar 
to the one found for the dilaton spurion breaking. Indeed, it corresponds to
{\it two} mutually non-local states condensing simultaneously 
(the $(0,1)$ and $(1,1)$ states). 
\begin{figure}
\centerline{
\hbox{\epsfxsize=6cm\epsfbox{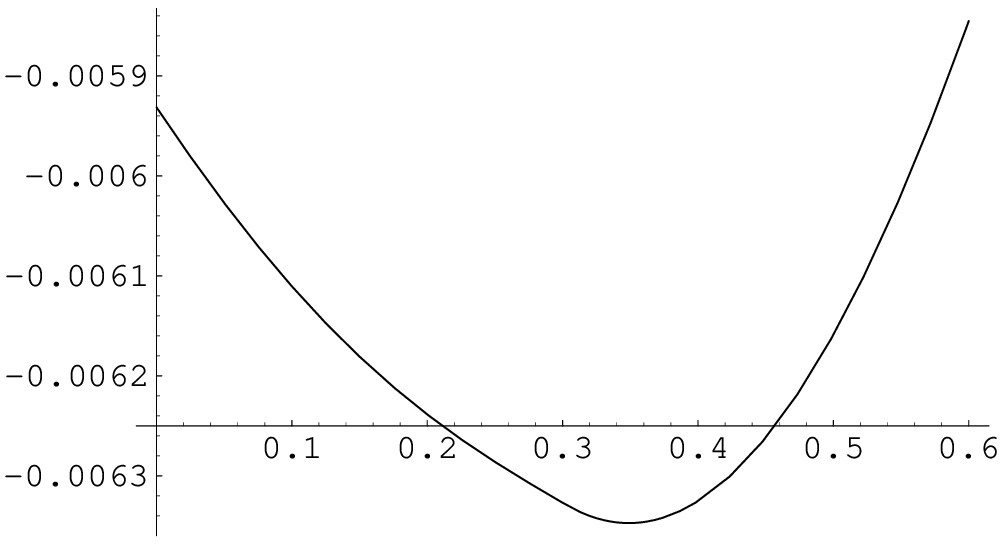}}\qquad
\hbox{\epsfxsize=6cm\epsfbox{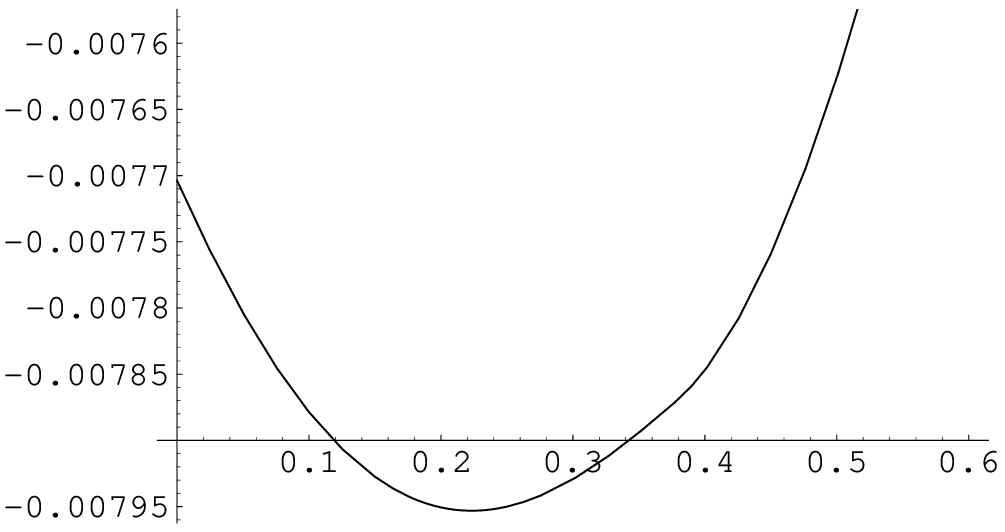}} }
\caption[]{Effective potential along the real $u$-axis for $m=0.1$ and 
$f_m=0.85$ (left) and $f_m=0.92$ (right).}
\figlabel\smooth
\end{figure}
As $f_m$ is still increased, the minimum 
moves smoothly along the real $u$-axis to the origin. At a certain point, 
the $(2,1)$ condensate is different from zero at the minimum and we have again 
a phase with {\it three} condensates, like the one we found for $m=0$.
 There are then 
two smoothly connected phases but with a different number of condensates 
at the minimum. This is shown in \fig\smooth, where the effective potential is 
plotted on the real $u$ axis for $m=0.1$ and $f_m=0.85$, $0.92$. In the first 
case the minimum has $\rho=0$ for the $(2,1)$ condensate, while $\rho \not= 0$ 
in the second case. In fact, the $(2,1)$ condensate can 
be considered as a continuous order parameter for this transition. 
The minimum moves smoothly to the origin 
along the real $u$ axis as $f_m$ is increased.    

The ``window" in which we have the phase with a condensate of
two mutually non-local states becomes bigger as $m$ is increased. For 
a certain 
value of $m$ this is the only non-local phase we find, as the 
singularity corresponding 
to the $(2,1)$ state moves away from the origin of the $u$-plane. The new 
phase with three mutually non-local states dissapears, and the situation is 
essentially similar to the one we found for the dilaton spurion induced 
supersymmetry breaking. The maximum allowed value for $f_m$ also decreases 
as we approach the Argyres-Douglas point, and $m=m_c$ is also a critical 
value for the mass in this model. The vacuum structure 
completely changes for $m>m_c$. What we find 
there is a single minimum on the real $u$ axis 
corresponding to a {\it quark} condensate. At this minimum there is 
a nonzero VEV 
for ${\tilde q}q$, as expected.  
\begin{figure}
\epsfxsize=6cm
\centerline{\epsfbox{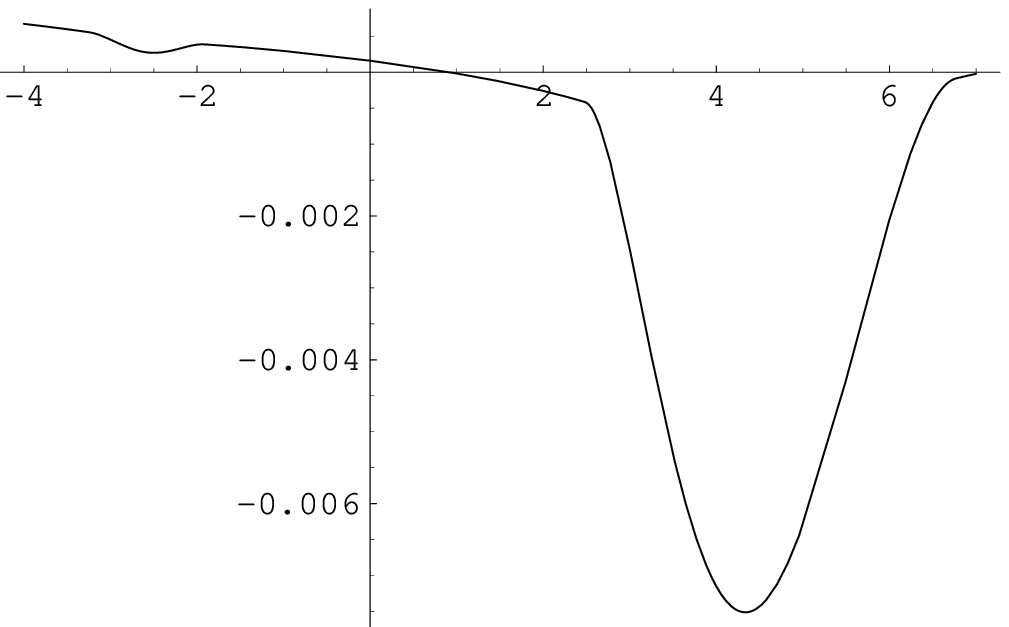} }
\caption[]{Effective potential along the real $u$ axis for $m=2$, $f_m =0.56$.}
\figlabel\quarkcond
\end{figure}
The $(2,1)$ dyon condensate is also different from zero, but smaller 
than the quark 
condensate in one order of magnitude. Finally, the monopole hardly condenses 
and in fact the cosmological constant is smooth at the monopole singularity. 
It is interesting to notice that, in the decoupling limit, the vacuum moves 
away to infinity together with the quark singularity. We have then a ``runaway 
vacua"-like phenomenon. This is expected, as the terms that lift the vacuum 
degeneracy of the $N=2$ supersymmetric theory in the bare Lagrangian are 
off-diagonal mass terms for the squarks, 
and when the quark hypermutliplet is decoupled they do not longer affect the 
$N=2$ vacuum structure. The whole $u$-plane of vacua is recovered.  
 
The situation is illustrated in \fig\quarkcond, where the effective potential 
is plotted along the real $u$-axis for $m=2$, $f_m=0.56$. The absolute 
minimum occurs around the quark singularity at $u \sim 4.2$. There is also a 
 relative minimum around the $(2,1)$ dyon singularity $u \sim -2.6$. 

\begin{figure}
\epsfxsize=5cm
\centerline{\epsfbox{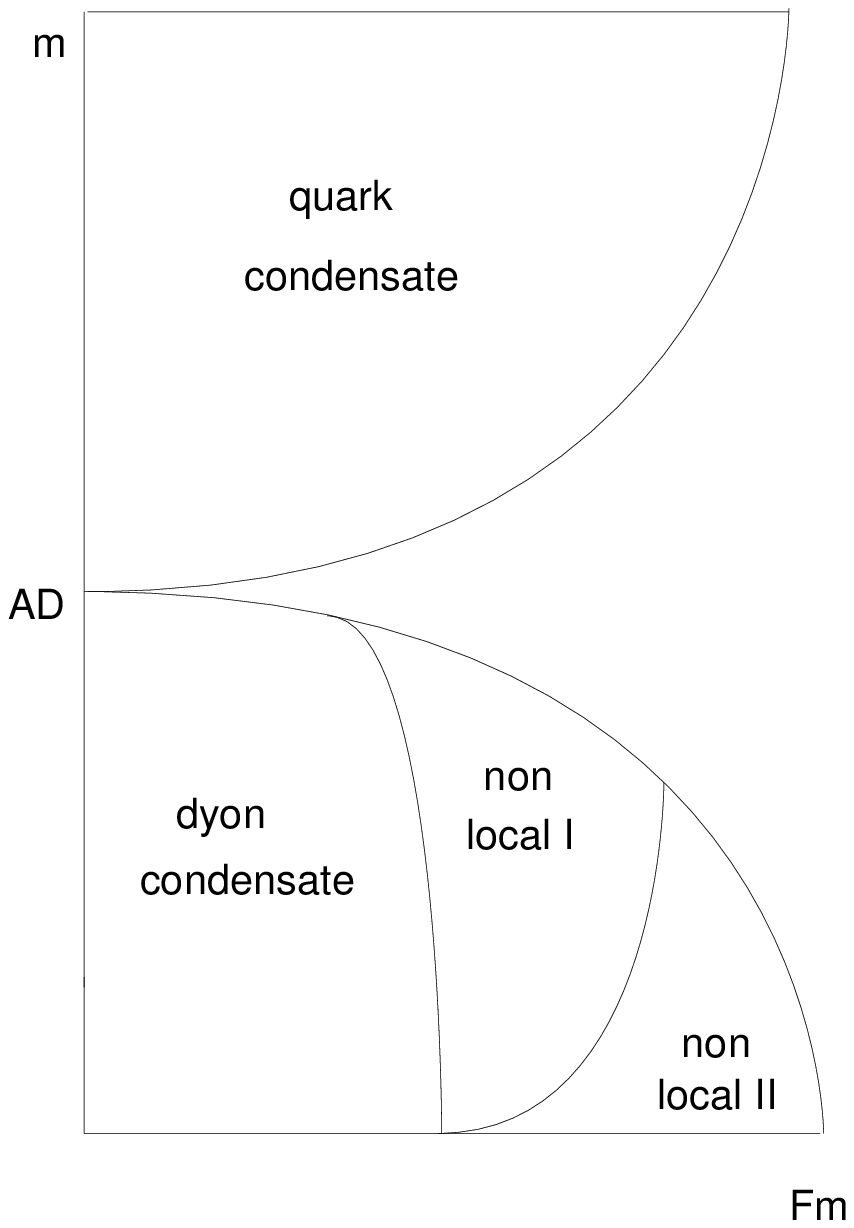} }
\caption[]{Phase structure of the $N_f=1$ theory for $m$, $f_m$ (denoted as 
Fm).}
\figlabel\diagramados
\end{figure}

The phases and vacuum structure when we turn on $m$, $f_m$, are summarized in 
\fig\diagramados. Again there are two different behaviours, depending 
on wether $m<m_c$ or $m>m_c$. When $m<m_c$, $m \not=0$, there are three
different phases for small values of $m$. The first one has two minima with 
dyon condensation and is separated from the second one 
(denoted by ``non local I") by a first order 
phase transition. This non-local phase exhibits simultaneous condensation of 
the $(0,1)$ and $(1,1)$ states, and is smoothly connected to a third phase 
(``non-local II" in the diagram) where there is also a condensate of the 
$(2,1)$ state.  
As $m$ increases, the third phase disappears, and as $m$ approaches $m_c$ 
from below the second phase also does (as in the dilaton spurion case). 
When $m>m_c$, the theory is in a different 
phase, and the vacuum corresponds to an elementary quark condensate.

\section{$N_f=2$ Vacuum Structure}
\setcounter{equation}{0}

In this section we analyse the vacuum structure of softly
broken $N=2$ QCD with two massive quark hypermultiplets.
We focus on the new physical phenomena arising because
of the non abelian flavor group.
In the first subsection the relevant symmetries are 
presented and the vacuum equations are solved for general
spurions. 
In the second subsection we study the case of breaking only with
dilaton spurion. In the next two subsections we discuss the 
phase transitions produced by the combination of mass and dilaton
spurions.

\subsection{Symmetries and Monopole Condensates}

With $N=2$ supersymmetry, the superpotential links the flavor rotations 
of the left handed quarks and antiquarks. For $SU(2)$ gauge group,
since the fundamental representation is pseudoreal, we can arrange the
$N_f$ flavors in a vectorial representation of $SO(2N_f)$.

For $N_f=2$ massless hypermultiplets, the global flavor symmetry 
group is $SO(4)$.  
Let us study more closely how the bare quark masses explicitly break 
 the flavor group. The mass term 
can be written 
with the $4 \times 4$ antisymmetric matrix $V^{rs} = Q^r Q^s$
 in the ${\bf 6}$ representation of $so(4)$. 
In terms of its 
selfdual and anti-sefldual parts, $V^{rs}= V^{rs}_{+} + V^{rs}_{-}$, 
the flavor group representation is $({\bf 3, 1}) \oplus 
({\bf 1, 3})$ of $so(4)= su(2)_{-} \oplus su(2)_{+}$,
 respectively. The mass term turns out to be
\be
\sum_{f=1}^2 m_f {\tilde Q}^f Q^f = 2 i m_{+} V_{-}^{13} 
+2 i m_{-} V_{+}^{13}.
\ee
$m_{\mp}=m_1 \mp m_2$ are the breaking parameters of the flavor subgroups 
$SU(2)_{\mp}$, respectively. 
Such variables are more adequate for the analysis
of the $N_f=2$ massive theory. For $m_{\mp}=0$,
the moduli space of the Coulomb branch of the theory 
has a singularity at $ u = \pm 1$, where a doublet
of magnetic BPS states in the $({\bf 2, 1})$ or $({\bf 1, 2})$ flavor
representations, respectively, become massless. 
>From these points a Higgs-confining branch emerges with one real 
modulus given by $<V_{\pm}> \not= 0$ and $<V_{\mp}> = 0$, for
$u=\pm 1$, respectively.  

\figalign{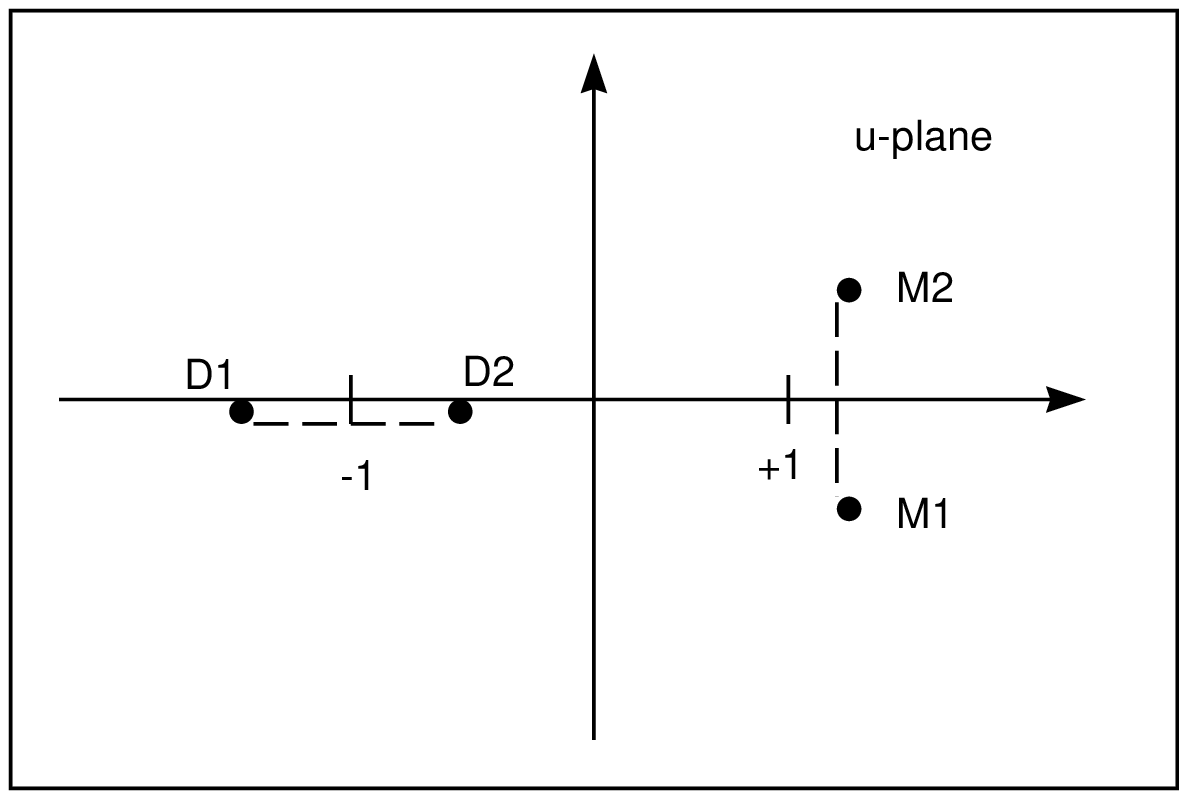}{\usings}{The singularities of the $u$-plane
for real bare masses $m_{\pm}$.}
{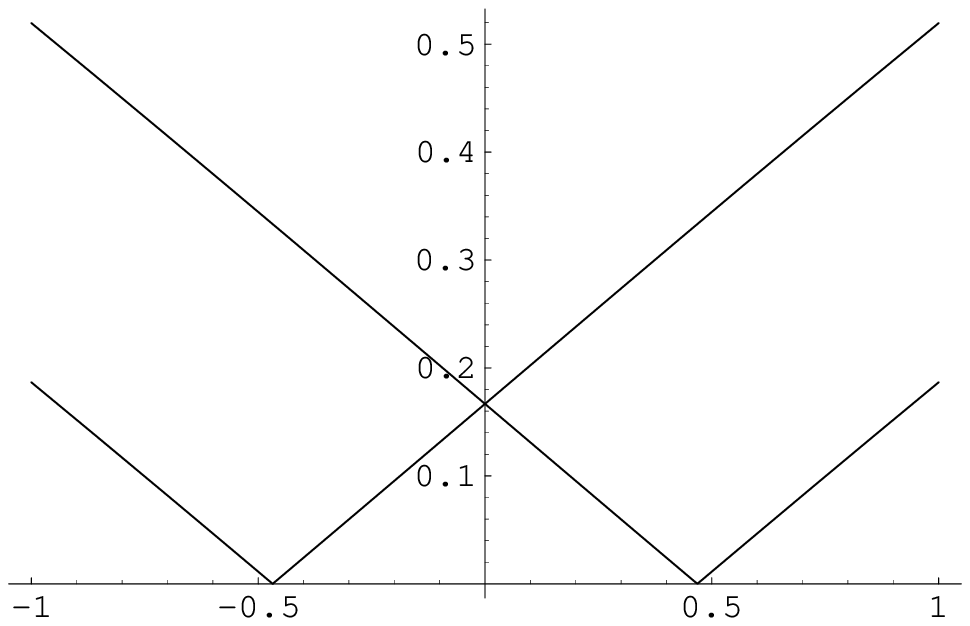}{\monomass}{Plot of the supersymmetric BPS masses of 
the two light monopoles M1 and M2 along the path joining the complex 
conjugate splitted singularities for $m_{\pm} =1/3$.}
If we turn on 
the bare parameter $m_{\mp}$, the singularity at $u= \pm 1$ splits 
in two singularities with a single massless state, showing 
the explicit breaking of the $SU(2)_{\mp}$ flavor subgroup
in the bare Lagrangian (see \fig{\usings}). 
The BPS masses of the two light
magnetic hypermultiplets also split because of the 
different baryon numbers of the
unbroken $U(1)_{\mp} \subset SU(2)_{\mp}$ symmetry 
(see \fig{\monomass}).  
For $N_f=1$ we could make a local inhomogeneous 
duality transformation on a given vanishing cycle
such that it does not enclose any simple pole of 
$\lambda_{SW}$, giving baryon number equal to zero
to the associated BPS state. But for $N_f=2$
there are two mutually local singularities.
We can choose one of them to have baryon numbers equal to zero,
but that fixes unambiguously the baryon numbers of the 
other one.
It is the quantity $S^{\mp}_2 - S^{\mp}_1$ 
which has a duality invariant meaning and 
carries all the information we need about the baryon numbers. 
For quark baryon numbers normalized to one, we have
$S^{\mp}_2 - S^{\mp}_1 = 1$ and $S^{\pm}_2 - S^{\pm}_1 =0$ for
the region around $u=\pm 1$, respectively. 

Then, for $N_f=2$ massive softly broken $N=2$ QCD, we have two well 
differentiated regions: the monopole region near $u=1$ and
the dyon region near $u=-1$.
In these regions, the relevant degrees of freedom at low 
energies are an $U(1)_G$ abelian vector multiplet and two 
light charged hypermultiplets. The soft $N=2$ supersymmetry
breaking terms generate a non trivial effective potential
for each region. The analysis   
becomes more clear if we capture the symmetries
broken by the small perturbations. We define the $2 \times 2$
matrix 
\be
\Phi^{ia} = \left( \matrix{h_1 & {\bar {\tilde h}}_1 \cr
h_2 & {\bar {\tilde h}}_2 } \right)
\ee
in the $({\bf 2}, {\bf 2}, 0, 1)$ representation of 
$SU(2)_{\mp} \times SU(2)_R \times U(1)_R \times U(1)_G$,  
for the monopole or dyon region, respectively. 
The action of the global symmetries is
\be 
\Phi^{ia} \rightarrow  g_{\mp}^{ij} \Phi^{jb} (g^{\dagger}_R)^{ba}.
\ee
The supersymmetry breaking parameters, as auxiliary terms of the $N=2$ 
spurion fields, can be thought as frozen vectors in the adjoint 
of $SU(2)_R$. They can be written as a unitary $2 \times 2$ matrix:
\be
{\bf F}^A = \left( \matrix{D^A & {\sqrt 2}F^A \cr 
{\sqrt 2}\, {\overline F}^A & -D^A } \right) = 
{\sqrt 2} {\rm Re}[F^A] \sigma^1 - {\sqrt 2} {\rm Im}[F^A] \sigma^2
+ D^A \sigma^3,
\ee
where $A = 0, +, -$.

The supersymmetric BPS masses of the two light hypermultiplets can be 
thought as frozen vectors in the adjoint of the $SU(2)_{\mp}$:
\be
{\bf M} = \left( \matrix{|{\sqrt 2} a + S^f_1 m_f| & 0 \cr
0 & |{\sqrt 2} a + S^f_2 m_f| } \right).
\ee

The effective potential is 
\bea
V_{{\rm eff}} &=&{1 \over 4} B_{AB} {\rm Tr} \{ {\bf F^A} {\bf F^B} \} 
+{1 \over 2 b_{aa}} \left( 2 {\rm Tr} \{(\Phi \Phi^{\dagger})^2\}
- ({\rm Tr} \{\Phi \Phi^{\dagger} \})^2 \right)
\nonumber
\\
&+&{\rm Tr} \{ {\bf M}^2 \Phi \Phi^{\dagger} \}
+ {1 \over b_{aa}} {\rm Tr} \{ \Phi (b_{aA} {\bf F}^A) \Phi^{\dagger} \}
\nonumber
\\
&-& \sum_{i=1}^2 S^f_i \Phi^{ia} {\bf F}_f^{ab} (\Phi^{\dagger})^{bi}.
\label{effpot}
\eea 
The first term corresponds to the cosmological constant,
where each of the six factors
$B_{AB} = ({b_{aA} b_{aa}^{-1} b_{aB}} - b_{AB})$
is duality invariant (I). 
The second term is $SU(2)_{\mp} \times SU(2)_R$ invariant.
The third term breaks $SU(2)_{\mp}$ explicitly and is
$SU(2)_R$ invariant.
The fourth term breaks $SU(2)_R$ explicitly and is
$SU(2)_{\mp}$ invariant.
The fifth term breaks $SU(2)_{\mp}$ and $SU(2)_R$
simultaneously. In the bare Lagrangian, $m_f$ and/or $F_f$ different
from zero break $SO(4) \rightarrow SO(2)_1 \times SO(2)_2$. But
at the level of the vacuum structure, as the monopoles (dyons)
 do not carry the quantum numbers
of $SU(2)_{+(-)}$, the effect at low energy 
of the $SU(2)_{\mp}$ breakings is only seen 
on the monopoles and dyons effective potentials, respectively.
Finally, we observe that
for each local effective potential, the $SU(2)_{\mp}$ breaking terms
are still invariant under
an abelian subgroup $U(1)_{\mp} \subset SU(2)_{\mp}$. 
Also, if we only break supersymmetry with the dilaton spurion, there is 
an unbroken $U(1) \subset SU(2)_R$ given by the direction 
of the vector ${\bf F}_0$.

Next we focus on the vacuum structure. We first solve for the 
minimum of the effective potential in terms of the monopole 
condensates for a given vacuum $u$. 
As in the previous section, we consider the case of supersymmetry breaking 
purely by $F$-terms, i.e., we put $D^A = 0$ for all spurions, 
being the difference only an $SU(2)_R$ rotation of the monopole
condensate solutions.
The equations are
\bea
{\partial V_{{\rm eff}} \over \partial {\bar h}_i} = 
{1 \over b_{aa}} (|h|^2 - |{\tilde h}|^2) h_i 
+{2 \over b_{aa}} ({\tilde h} h) {\bar {\tilde h}}_i 
\nonumber
\\
+|{\sqrt 2} a_i|^2 h_i
+ {{\sqrt 2} \over b_{aa}}{\overline F}_i {\bar {\tilde h}}_i = 0,
\label{eqh}
\eea
\bea
{\partial V_{{\rm eff}} \over \partial {\bar {\tilde h}}_i} = 
{-1 \over b_{aa}} (|h|^2 - |{\tilde h}|^2) {\tilde h}_i 
+{2 \over b_{aa}} ({\tilde h} h) {\bar {\tilde h}}_i 
\nonumber
\\
+|{\sqrt 2} a_i|^2 {\tilde h}_i
+ {{\sqrt 2} \over b_{aa}} {\overline F}_i {\bar h}_i = 0,
\label{eqantih}
\eea
where, to simplify notation, we have defined
$$
|h|^2 = \sum_{i=1}^2 |h_i|^2, \quad\quad
{\tilde h} h = \sum_{i=1}^2 {\tilde h}_i h_i,
$$
\be
a_i = a + S^f_i {m_f \over {\sqrt 2}}, \quad\quad
F_i = b_{aA}F^A - b_{aa}S^f_i F_f.
\ee

If we consider the combination
\be
{\bar h}_i {\partial V_{{\rm eff}} \over \partial {\bar h}_i} 
-  {\bar {\tilde h}}_i {\partial V_{{\rm eff}} \over 
\partial {\bar {\tilde h}}_i} =0
\ee
we obtain a linear sistem of homogeneous equations for the variable
$(|h_i|^2 - |{\tilde h}_i|^2)$ with determinant different from zero.
This implies that $|h_i| = |{\tilde h}_i|$, for each $i=1, 2$, 
a consequence of
breaking only with $F$-terms. We make the parametrization
$h_i = \rho_i e^{i\b_i}$ and ${\tilde h}_i = \rho_i e^{i {\tilde \b}_i}$,
to obtain the vacuum equations
\be
\rho_i \left( \sum_{j=1}^2 (\rho_j^2 e^{i (\b_j + {\tilde \b}_j) }) +
 b_{aa}|a_i|^2 e^{i (\b_i + {\tilde \b}_i) } 
+ {{\overline F}_i \over {\sqrt 2}} \right) = 0.
\label{vaceq}
\ee

If $\rho_i \not= 0$ for all $i=1,2$, subtracting both equations 
 (\ref{vaceq}) we obtain the compatibility equation
\be
|a_2|^2 e^{i (\b_2 +{\tilde \b}_2) } - |a_1|^2 e^{i (\b_1 +{\tilde \b}_1) } 
= { {\overline F}_{\mp} \over {\sqrt 2}}
\label{compatibility}
\ee
 near $u= \pm 1$, respectively. 
If (\ref{compatibility}) is not satisfied, 
then we cannot have simultaneous condensation of the mutually
local monopoles. The solution is
\bea
\rho_i &=& - b_{aa}|a_i|^2  
+ {|{\overline F}_i| \over {\sqrt 2}},
\label{rhocondensate}
\\
\b_i + {\tilde \b_i} &=&{\rm Arg}[ {\overline F}_i] + \pi,
\label{monophase}
\\
\rho_j &=& 0 \quad\quad (j \not= i).
\eea
If we introduce the monopole condensate into the effective potential,
we obtain the $u$ dependent function
\be
V_{{\rm eff}}(u) = B_{AB}(u) {\overline F}^A F^B 
- {2 \over b_{aa}(u)}\rho^4(u).
\ee
with $\rho = {\rm Max}[ \rho_1, \rho_2]$, the bigger condensate which
minimizes the vacuum energy. 
As the monopoles have different baryon numbers,
the potential is anisotropic. There are different 
prefered directions for the vector $(\rho_1, \rho_2)$
depending on the region of the moduli space. 
At the transition point $\rho_1 = \rho_2$ 
there is a flip in the VEV direction
which produces a
discontinuous first derivative of $V_{{\rm eff}}(u)$.
This cusp in the effective potential 
shows a first order phase transition associated to
the level crossing of the $\rho_1$ and $\rho_2$ condensates:
on the cusp line, the configuration minimizing the vacuum energy
changes from one direction to the other, {\it i.e.},
from the energy level given by the $\rho_1$ VEV to the one
given by the $\rho_2$.
We expect to smooth this
cusp if the next to leading contributions to the effective potential 
are included.

\subsection{Breaking with only Dilaton Spurion}

\begin{figure}
\centerline{
\hbox{\epsfxsize=6cm\epsfbox{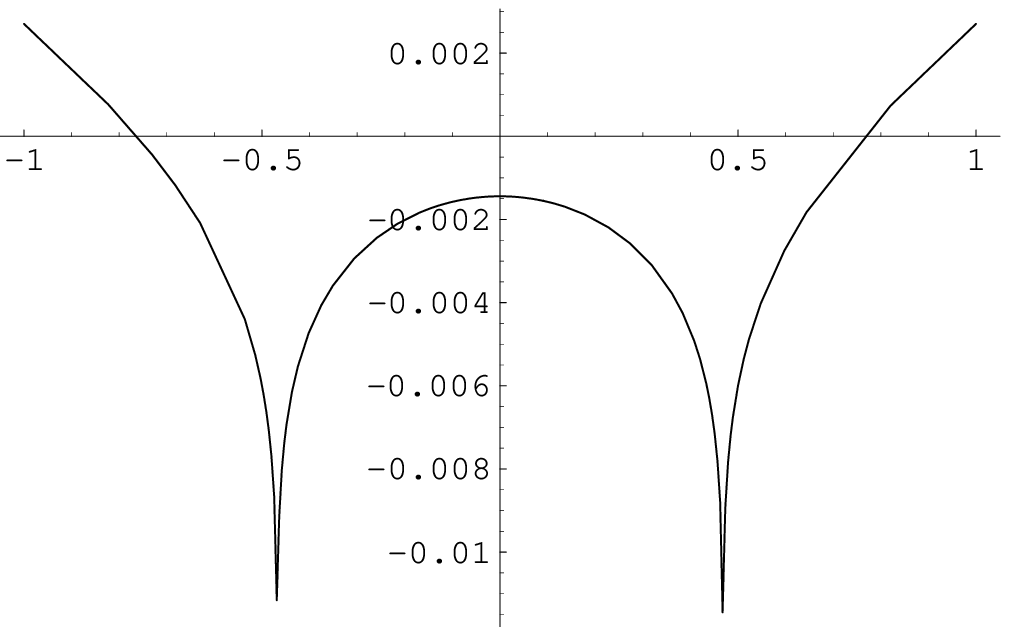}}\qquad
\hbox{\epsfxsize=6cm\epsfbox{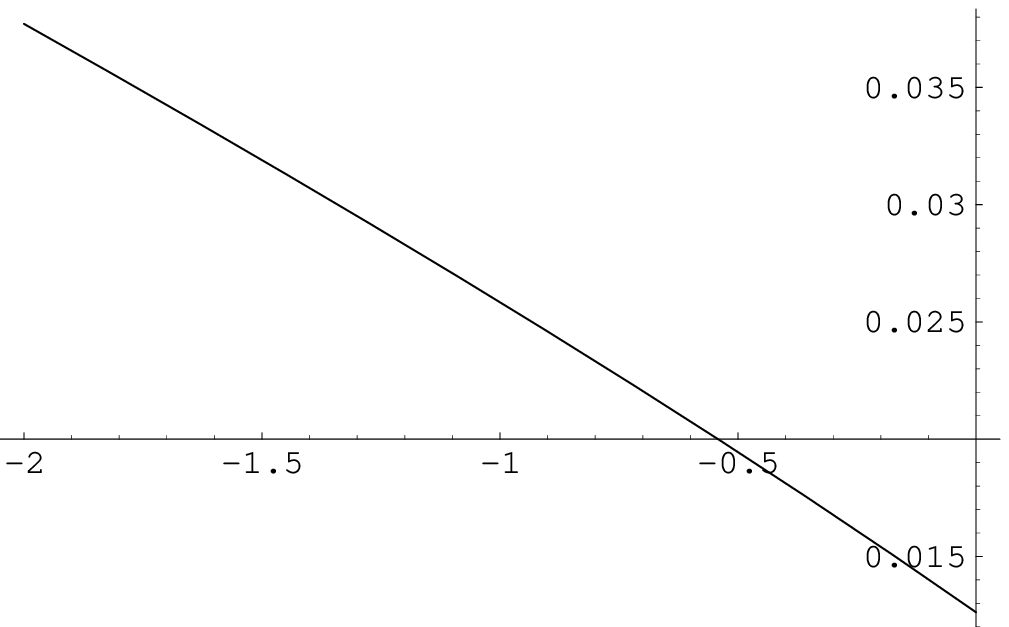}} }
\caption[]{Plot of the $|F_0|^2$ cosmological constant factor $B_{00}(u)$
through the path which joins the two mutually
local singularities in the monopole region (left) and in the 
dyon region (right) for $m_{\pm}=1/3$.}
\figlabel\cosmdilaton
\end{figure}

In this subsection we focus on the vacuum structure originated by
the supersymmetry breaking only with  $F_0 \not= 0$, putting 
mass spurions $F_f = 0$. The massless case was studied in 
\cite{soft}. In the numerical analysis we choose the normalization
 $\Lambda^2_2 =8$.

The order parameter which makes possible 
the monopole condensate is the coupling
$b_{a0}$. As in the massless case, it is practically zero near 
the dyon region, with a cosmological constant already
smooth at $u\sim -1$ (see \fig{cosmdilaton}, right). 
The dilaton parameter explicitly breaks  
the ${\bf Z}^R_2 \subset U(1)_R$ symmetry and 
the minimum is located at the monopole region 
$u \sim 1$.

For real bare quark masses, 
the supersymmetry breaking order parameter $b_{a0}$ and the cosmological
constant are symmetric with respect to the CP action 
$u \rightarrow {\bar u}$ (see \fig{\cosmdilaton}, left).
When we include the monopole condensate contribution to the 
effective potential we obtain two smooth minima near the two 
splitted monopole singularities. These
two minima are energetically degenerate, located at complex 
conjugate points in the $u$-plane. 
There is an spontaneous symmetry breaking of the CP symmetry.
The states condensing at these vacua 
have opposite physical electric charges $q(u)= 
\theta_{{\rm eff}}(u) / \pi \not= 0$, induced by 
Witten's effect \cite{dyon}. 

If we give complex phases $m_{\mp} \rightarrow e^{i\alpha_{\mp}}m_{\mp}$
to the bare masses, the $u$ singularities rotate around $u= \pm 1$,
 respectively. If their phase is the same, we can perform
an $U(1)_R$ anomalous transformation to 
absorb the common global phase into the bare $\theta$ angle. 
This corresponds to a rotation of the $u$-plane \cite{hsuangle}, and we 
obtain the same physical situation. 
But if the bare quark masses have a relative complex phase,
the CP symmetry is lost and we only have one absolute
minimum, which is always located at the singularity with the biggest value
for $|u|$.

The cusp line is on the points
of the $u$-plane where the two monopoles have the same mass,
which for real bare quark masses is on the real $u$-axis
(see \fig{\monomass}).
\figalign
{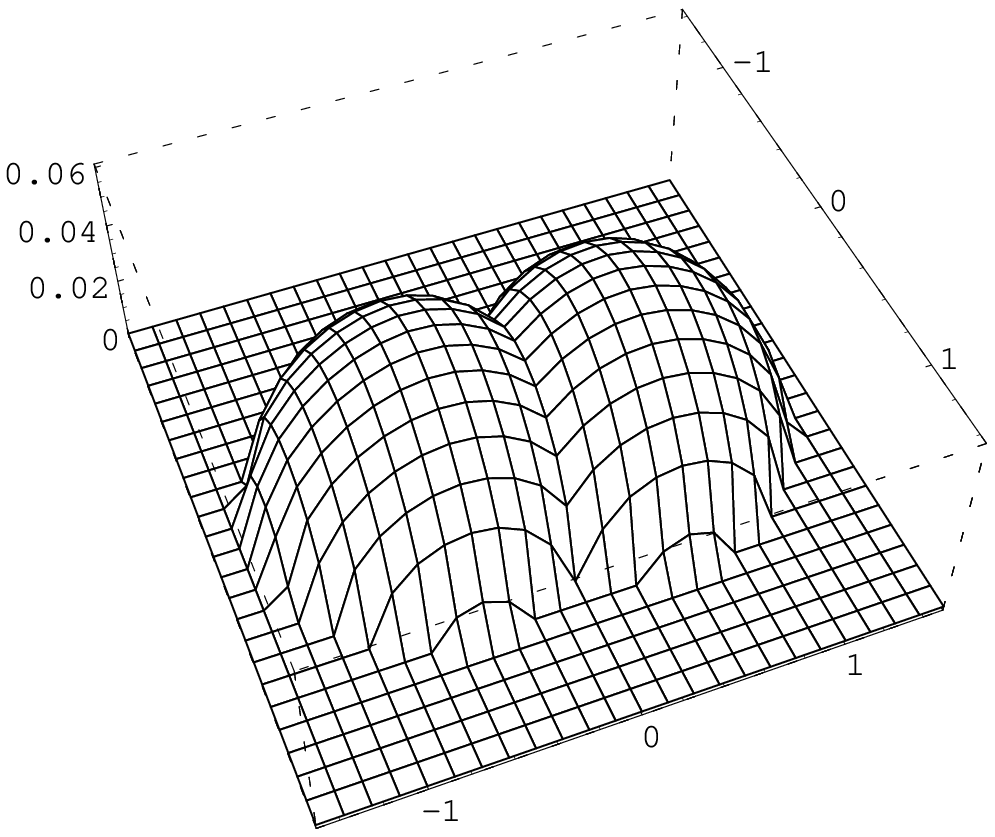}{\mrhodilaton}
{The monopole condensate $\rho= {\rm Max}[\rho_1, \rho_2]$
for $f_0=1/10$ and $m_{\pm}=1/3$.}
{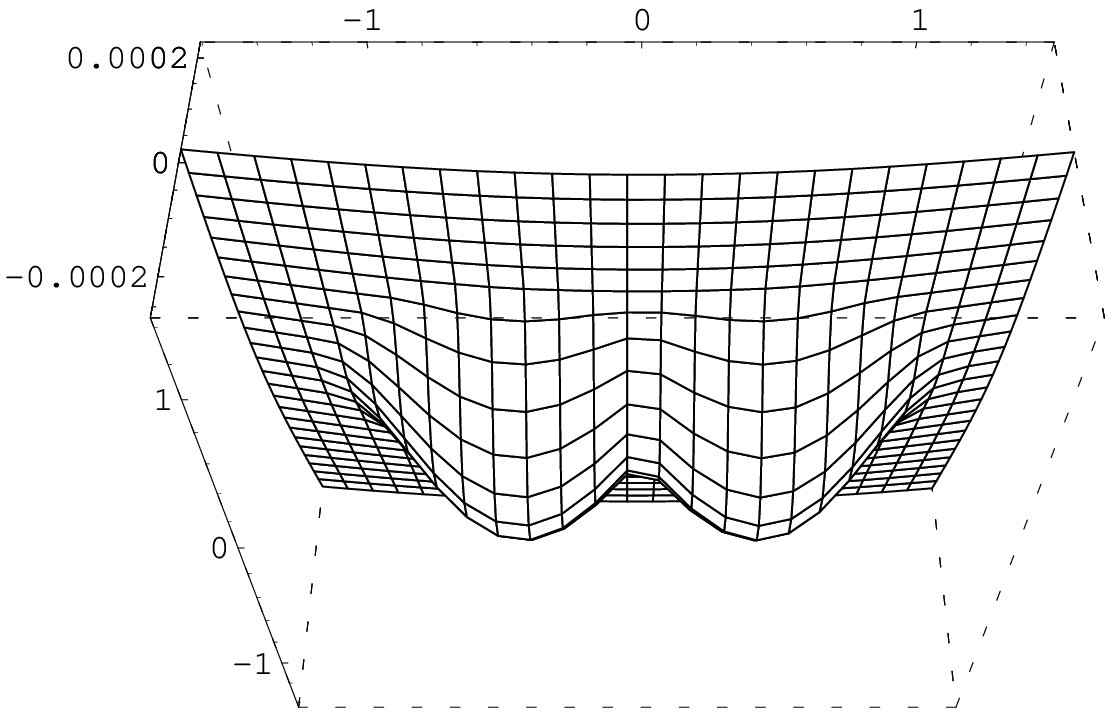}{\mveffdilaton}
{The effective potential in the monopole region for $f_0=1/10$
 and $m_{\pm}=1/3$.} 
The cusp appears when the two  
condensates start to overlap, for values of the dilaton spurion
bigger than some critical value $F_0^{(c)}(m_{-})$ 
(See \fig{\mrhodilaton} and \fig{\mveffdilaton}).

\subsection{Breaking with Mass Spurions: Phase Transitions
between Mutually Local Minima}

When we turn on the mass spurion supersymmetry breaking parameters, a rich 
phase structure emerges. 
The first thing to notice is that when $|F_{-}|$ 
or $|F_{+}|$ are increased, the absolute minimum tends to be 
 in the monopole or dyon region, respectively.
We will study these phenomena in the next subsection.
In this subsection we shall consider that there is still
a value of $F_0$ large enough for
the absolute minimum to be in the $u \sim 1$ region, 
and focus on the phase transitions between the minima 
associated to mutually local singularities.

\begin{figure}
\centerline{
\hbox{\epsfxsize=4cm\epsfbox{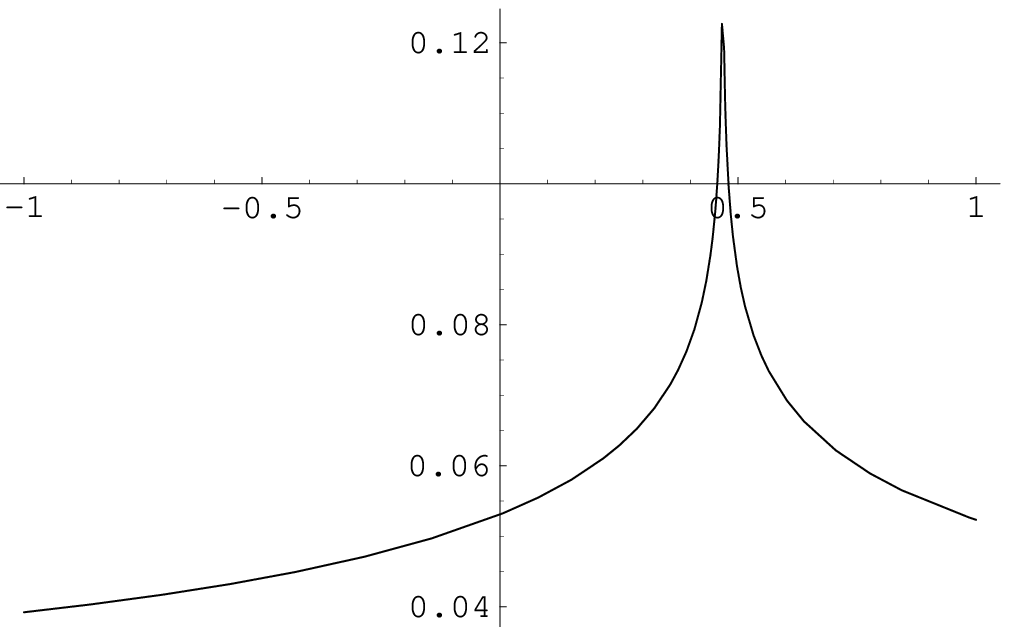}}\qquad
\hbox{\epsfxsize=4cm\epsfbox{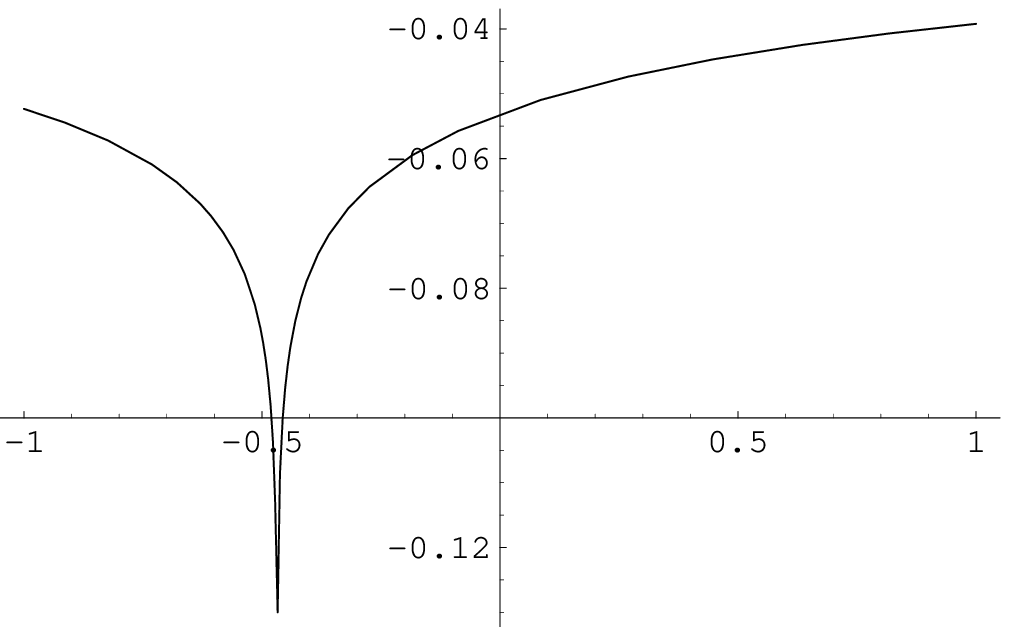}}\qquad
\hbox{\epsfxsize=4cm\epsfbox{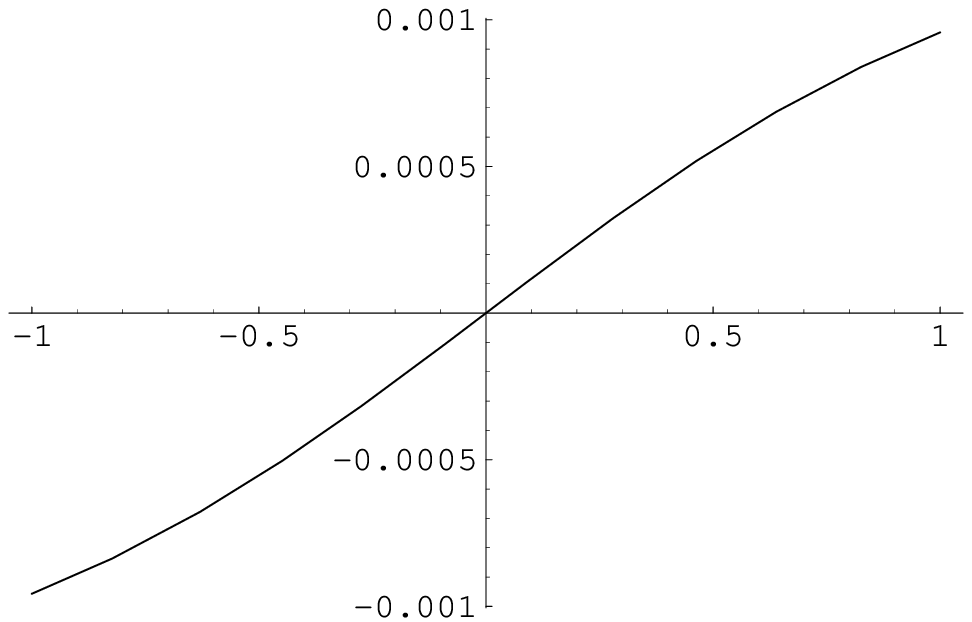}} }
\caption[]{
The supersymmetry breaking mass couplings $b_{a-}^{(1)}$ (left),
$b_{a-}^{(2)}$ (middle) and $b_{a+}$ (right) along the path
joining the complex conjugate monopole $u$-singularities for $m_{\pm}=1/3$.}
\figlabel\mbaA
\end{figure}

When we turn on the mass spurions, we
must be careful with the monopole condensates $\rho_i$. 
We have seen that the vacuum chooses the biggest local condensate
$\rho={\rm Max}[\rho_1, \rho_2]$.
If one breaks supersymmetry with only the dilaton spurion,  
the functions $\rho_i$ (\ref{rhocondensate}) 
go to zero on the $u_j$ ($j\not=i$), and
for each singularity,
the vacuum chooses always its corresponding local condensate.
We can take the massless
limit keeping $F_0$ different from zero,
and then we recover the situation described in
\cite{soft}. At the same time, the cusp disappears
as the flavor symmetry is restored.
But when we turn on the mass spurions, 
their contribution to the $\rho_i$ condensate
\footnote{To symplify, we stay on the monopole region, 
but the same analysis can be extended to the dyon region 
just by interchanging the $-$ and $+$ indices.} 
is proportional to 
$b_{a-}^{(i)}= b_{a+}F_+ +(b_{a-} - b_{aa}S_i^{-})F_-$.
This is the adequate coupling near the $u_i$ singularity,
as it corresponds to the right local variable there, 
$a_i = a +S_i^{-}m_{-} / {\sqrt 2}$. 
This coupling does not diverge at $u_i$.
\begin{figure}
\centerline{
\hbox{\epsfxsize=6cm\epsfbox{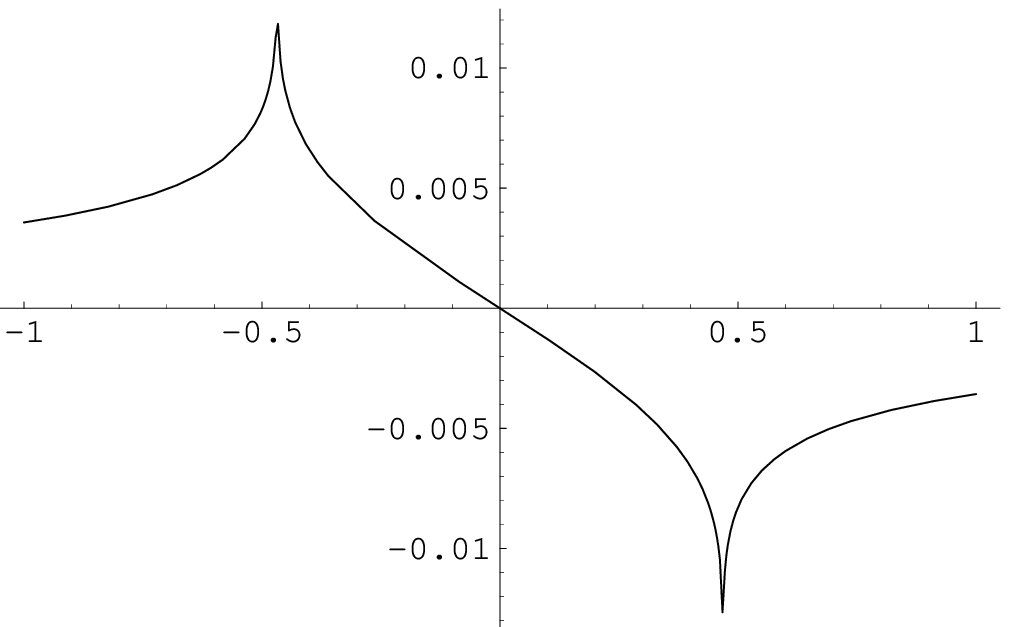}}\qquad
\hbox{\epsfxsize=6cm\epsfbox{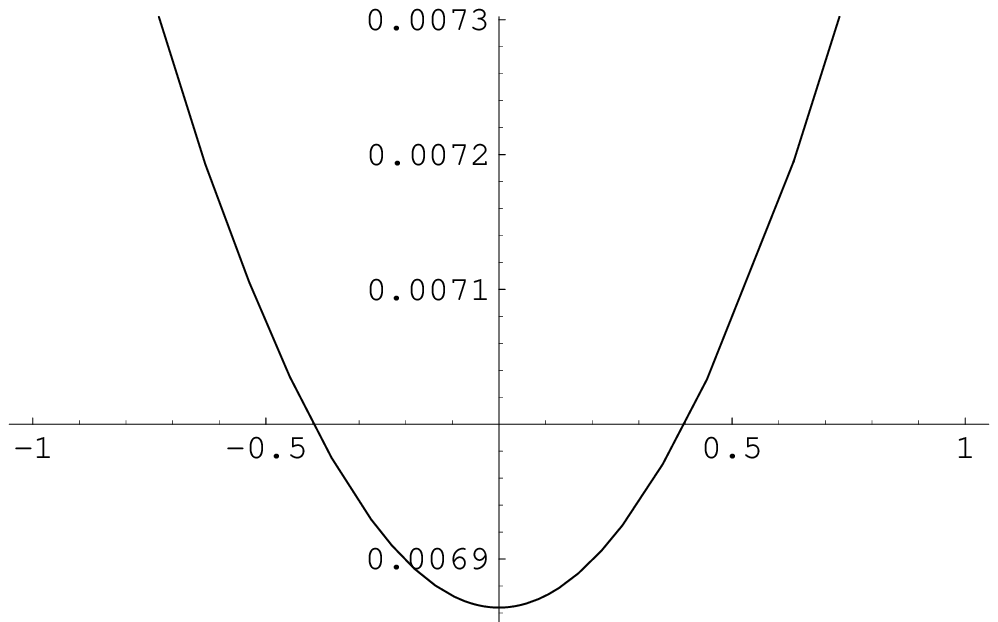}}}
\caption[]{ The cosmological constant factors $B_{0-}$ (left)
and $B_{++}$ (right) along the path joining the complex conjugate 
monopole singularities for $m_{\pm}=1/3$.}
\figlabel\mcosmpn
\end{figure}

But then it will inevitably diverge at the other mutually local
$u_j$ singularity ($j\not=i$) (see \fig{\mbaA}). 
To avoid this divergence, the $\rho_i$ condensate must
not attain the $u_j$ singularity, and 
this condition will give an upper bound for the mass spurions.
Let us consider the condensate $\rho_2$ in terms of the
$a_1$ variable, which
has the baryon numbers $(S^{+}_2, S^{-}_2) = (0, 1)$. We can
 redefine the monopole phases such that $F_-$ is real and positive, and 
then we expand in powers of $1/ (b_{aa}F_-)$:
\bea
\rho_2 &=& b_{aa}\left( F_- -{1 \over {\sqrt 2}} 
|{\sqrt 2} a_1 +m_-|^2 \right)
\nonumber 
\\
&-&{1 \over {\sqrt 2}}{\rm Re}[b_{a0}F_0 + +b_{a+}F_+ +b_{a-}^{(1)}F_-]
+ {\cal O}({1 \over b_{aa}F_-}),
\eea
which goes to infinity at $u_1$ for $F_- > |m_-|^2 /{\sqrt 2}$.
The $b_{a+}$ coupling, as the monopoles do not have 
$SU(2)_{+}$ quantum numbers, remains finite at both 
singularities (see \fig{\mbaA}, right). Actually, its contribution to 
the condensate is two orders of magnitude smaller than 
the one associated to the $b_{a-}^{(i)}$ couplings. 

Then, if we do not want the monopole and dyon condensates 
to explode at the other mutually local singularities, 
we must impose the upper bounds 
$F_{\pm}^{(Max)}= |m_{\pm}|^2 / {\sqrt 2}$ to the mass
supersymmetry breaking parameters.
In the massless limit, $m_f \rightarrow 0$, we must also 
send the mass spurions to zero. 
This indicates that for mass spurions bigger than these
bounds, the effective potential is unbounded 
from below. This instability of the vacuum is 
expected from the structure of the soft breaking terms
in the bare Lagrangian, as we already mentioned in section 2. 

\begin{figure}
\centerline{
\hbox{\epsfxsize=6cm\epsfbox{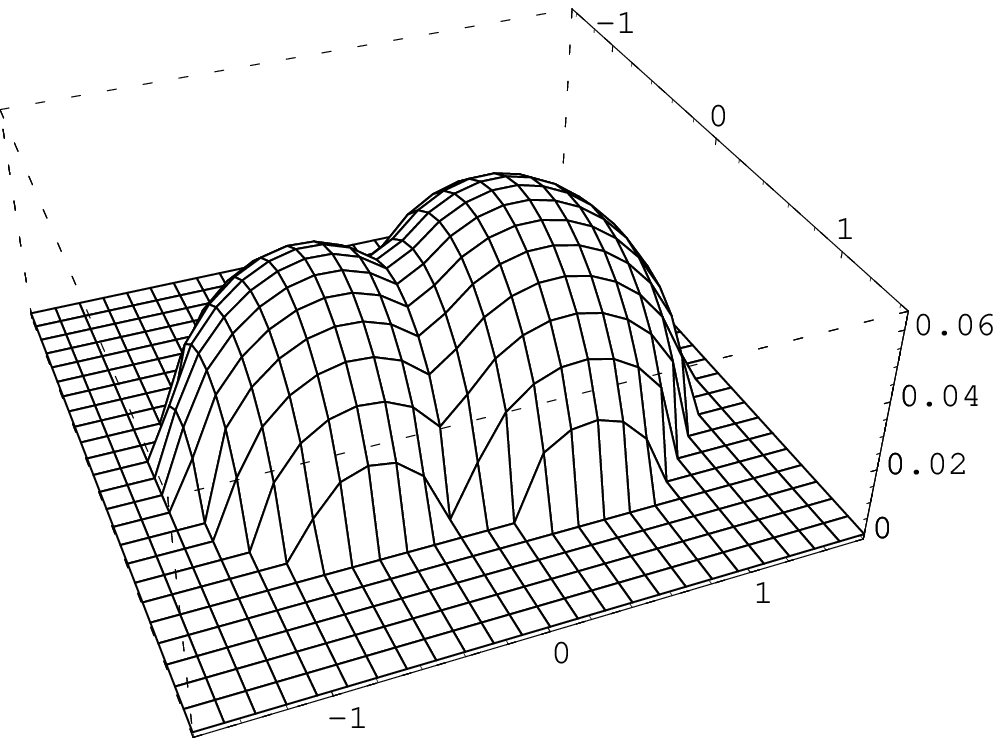}}\qquad
\hbox{\epsfxsize=6cm\epsfbox{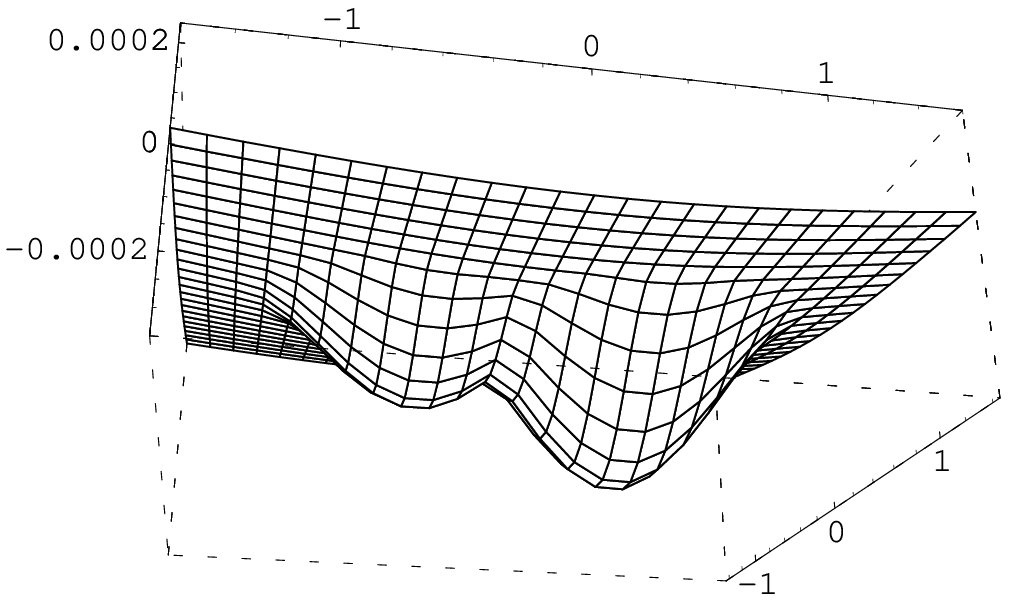}} }
\caption[]{The monopole condensate (left) and 
the effective potential (right) for $F_0=1/10$ and
$F_- = -{1 \over 5}F^{max}_-$.}
\figlabel\localtrans
\end{figure}

For real bare masses, the cosmological constant factors $B_{0\mp}(u)$ 
and the couplings $b_{a-}^{(i)}$ and $b_{a+}$
are antisymmetric with respect to 
the CP action (see \fig{\mcosmpn} (left) and \fig{\mbaA}). 
Then, for two generic spurions different from zero, 
there is a first order phase transition 
between local minima, as we change the sign of
their relative phase.
To be concrete, take $F_0$ and $F_-$ to be real numbers. When they have the 
same sign, the $\rho_1$ condensate is bigger than the $\rho_2$ 
condensate. Also, the effective potential has the absolute
minimum near $u_1$. 
When the spurions have opposite signs, we obtain 
the CP transformed situation, and now the absolute
minimum is near $u_2$
(see \fig{\localtrans}). 
The CP symmetry is explicitly broken in the effective potential.

\subsection{Breaking with Mass Spurions: Phase 
Transitions between Mutually Non-Local Minima}

When we decrease $|F_0|$ with respect to $|F_{\pm}|$,
the dyon region enters the game. For real bare masses,
the cosmological constant factor $B_{--}$ looks like the factor $B_{00}$ 
in the monopole and dyon regions (see \fig{\cosmdilaton}).
In the dyon region, $B_{++}$
is like $B_{--}$ in the monopole region and the supersymmetry 
breaking couplings $b_{a+}^{(i)}$ are of the same order of 
magnitude as the $b_{a-}^{(i)}$ couplings in the monopole region.
For $|F_+| > |F_-|$, if we decrease $|F_0|$, 
the effective potential in the monopole region is dominated  
by the smooth cosmological constant factor $B_{++}$
(see \fig{\mcosmpn}, left). There is a critical value of the 
dilaton spurion for which  
the absolute minimum jumps to the dyon region,  
and we have a first order phase transition between 
mutually non-local minima, which have different patters of
chiral symmetry breaking.
 
For $F_0 = 0$, the ${\bf Z}^R_2$ symmetry in the bare 
Lagrangian is only broken by the bare quark masses.
If we take complex conjugate masses, $m_1 = {\overline m}_2$, 
the M1 and M2 monopole singularities (see \fig{\usings}) rotate 
counterclockwise and all the singularities
of the $u$-plane are on the real axis. It is only in 
this extreme situation that
there is a complete symmetry between the monopole and dyon region,
if we also interchange the variables with indices
$+$ and $-$. From now on we will consider this symmetric 
situation.

For $|F_{-}| > |F_+|$, the absolute minimum is
in the monopole region. When their relative phase 
changes sign, there is a first order phase transition
between the two monopole minima, as the one plotted in \fig{\localtrans}.  
Across the lines $|F_{+}| = |F_{-}|$ there is a
first order phase transition between the mutually non-local minima.
These minima give different realizations of the chiral symmetry breaking 
pattern, and we have 
$SU(2)_- \times SU(2)_+ \rightarrow SU(2)_{\pm}$ for the monopole
and dyon regions, respectively.

\begin{figure}
\centerline{
\hbox{\epsfxsize=6cm\epsfbox{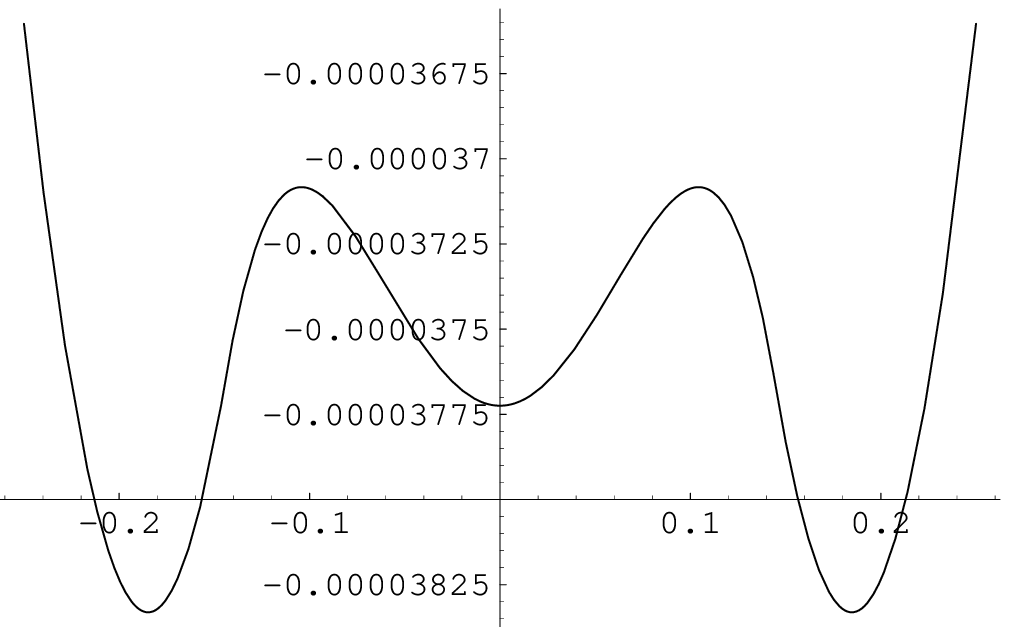}}\qquad
\hbox{\epsfxsize=6cm\epsfbox{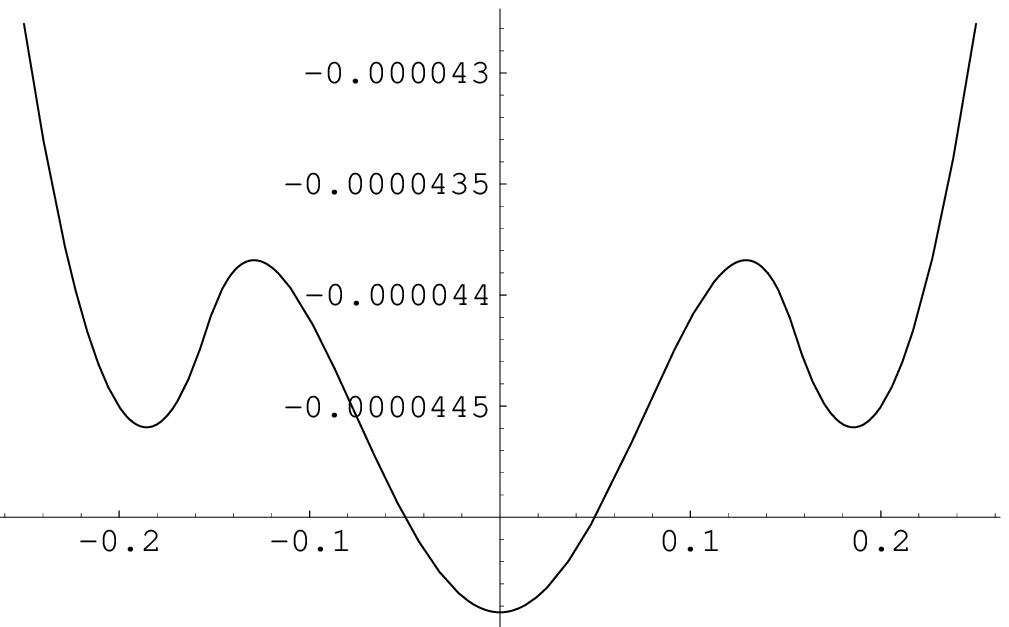}} }
\caption[]{The non-local first order phase transition for 
$m_1 = {\bar m_2}= {7 \over 10}(1+i)$ and 
$F_- = F_+ = 0.15 F_{\pm}^{Max}$.}
\figlabel\nonlocalveff
\end{figure}

If we simultaneously increase $|m_{+}|$ and $|m_-|$, the monopole
and dyon singularities which are nearest to the 
origin approach each other, and they meet at 
the superconformal  AD $(1,1)$ point at $u=0$ for $m_+=1$ 
and $m_- = i$ \cite{apsw}. But before that happens,
the monopole and dyon condensates begin to overlap.
For a critical value of the bare quark masses 
($|m_{\pm}| \sim 7/10$), this overlaping gives rise
to a new minimum at $u=0$, provided that
the supersymmetry breaking parameters are large enough 
(see \fig{\nonlocalveff}). 
Once again, we find a first order phase transition
similar to the ones considered in \cite{softdos}
and in the previous section.
Both states have opposite
physical electric charges at $u=0$, and we also observe 
in this case the possibility
of the formation of an electrically neutral ``bound state" at $u=0$.
We then see that these kinds of phenomena are generic
in the $N=2$ softly broken models.
The interesting thing of the $N_f=2$ case
is that this ``bound state" is made of mutually non-local
states with different flavor quantum numbers.
The monopole is in the flavor representation $({\bf 2, 1})$ and the 
dyon in the $({\bf 1, 2})$ of the $SU(2)_- \times SU(2)_+$ flavor 
group. The formation of this bound state could be 
a hint that the new
absolute minimum breaks $SU(2)_+ \times SU(2)_-$ to the diagonal 
$SU(2)_V$.

\begin{figure}
\epsfxsize=8cm
\centerline{\epsfbox{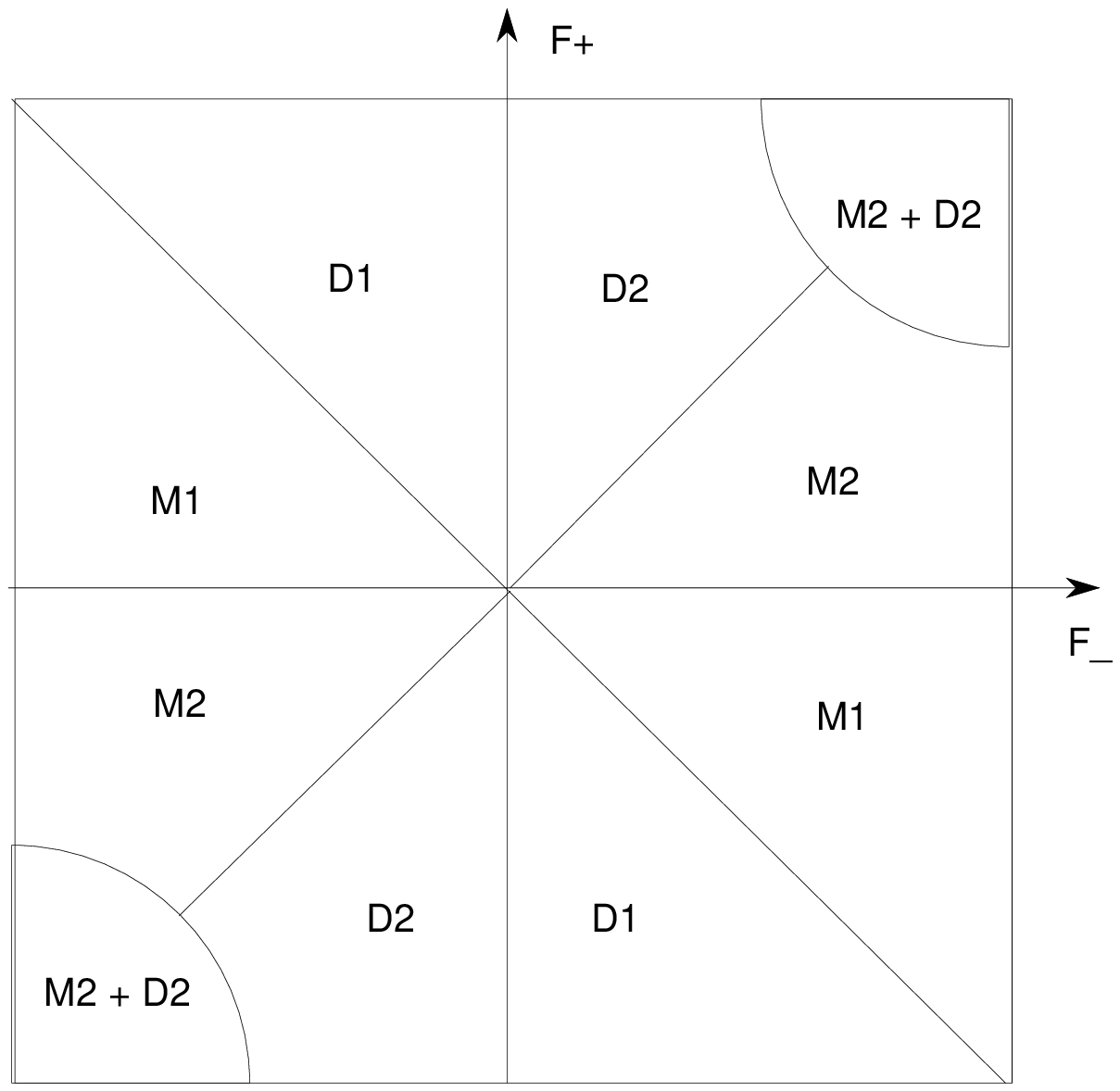} }
\caption[]{Phase diagram with $F_0=0$ and $m_1 = {\bar m_2}
\approx {7 \over 10}(1 +i)$
to have the M2 + D2 phase. M2 means a vacuum dominated by 
the condensate of the monopole nearest to $u=0$. D2 means the same thing, 
but for 
the dyon. M1 and D1 are the other more distant singularities.}
\figlabel{\phases}
\end{figure}
Finally, in \fig{\phases} we give the phase diagram of 
$N_f=2$ massive softly broken
$N=2$ QCD, for dilaton spurion $F_0 = 0$ and complex conjugate 
bare quark masses with modulus 
bigger than the critical value needed to see the 
 phase transition at $u=0$. The corners labeled by M2 + D2
 correspond to this new phase. To symplify, 
the mass spurions are taken to be real. 
There are first order phase transitions between mutually local minima
when their relative sign changes, {\it i.e.}, 
across the $F_{\mp}$ axis.
There are also first order phase transitions across the 
diagonals $|F_-|= |F_+|$, where the absolute minima
jumps between the monopole and dyon regions.
For given values of the bare masses, there are 
maximum allowed values for the mass spurions.
This is why the phase diagram is enclosed 
into an square box, in such a way  that none of 
the condensates reaches another singularity.

If we compute the squark condensates $<{\tilde q}_f q_f>$ given in
(\ref{squark}) at the minimum, for the massless case, we can check that 
$< V_+> \not=0$ and $<V_-> =0$. This is consistent with the spontaneous
symmetry breaking of the $SU(2)_-$ flavor subgroup due to the VEV of the 
monopole doublet. In the following section we give the pion Lagrangian
associated to this spontaneous symmetry breaking.

\section{The Pion Lagrangian}
\setcounter{equation}{0}

One of the main features of the Seiberg-Witten solution in \cite{swtwo} 
is that chiral symmetry breaking is naturally explained in terms of 
monopole condensation, once $N=2$ supersymmetry is broken down to $N=1$ 
\cite{swtwo} or to $N=0$ \cite{soft}. As we also have a low-energy effective 
Lagrangian describing the dynamics of the monopoles, one can try to 
obtain the ``pion" Lagrangian up to two derivatives 
for the Goldstone bosons associated to 
the chiral symmetry breaking. This provides an interesting information about 
the dependence of phenomenological parameters like the pion mass and $F_{\pi}$ 
in terms of the magnetic monopole description. Although the pattern of chiral 
symmetry breaking in the softly broken models is not completely 
QCD-like, certain aspects 
of this analysis could be of interest in more realistic models. 

When we soflty break supersymmetry down to $N=0$ with only a dilaton spurion  
in the $N_f=2$ theory with massless hypermultiplets, the flavor 
symmetry is not explicitly broken, and the minimum is located in the monopole 
region. Near  
the monopole singularity, the global symmetry of the model is $G=SU(2)_{-} 
\times 
U(1)$. The $U(1)\subset SU(2)_R$ corresponds to the unbroken subgroup
given by the direction of the $SU(2)_R$ vector, ${\bf F}_0$.
This will be our chiral symmetry group. For real and 
positive $F_0=f_0$ and $D_0=0$, the unbroken $U(1)$ can be 
parametrized by a $\theta$ 
angle as $g_1= {\rm exp}(-i \sigma_1 \theta)$. We parametrize $SU(2)_{-}$ 
in the standard way as 
$g_{-}= {\rm exp}(-i {\bf \theta} 
\cdot {\bf \sigma})$, where ${\bf \theta}=(\theta_1,\theta_2, \theta_3)$ 
and ${\bf \sigma}$ are the Pauli matrices. The monopole VEV spontaneously 
breaks $G=SU(2)_{-} \times U(1)$ down to an $H \simeq U(1)$ subgroup. To 
obtain the 
precise form of the symmetry breaking pattern, we choose 
a minimum with $h_1 \not= 0$, $h_2=0$. We then have
\be 
\Phi_0 = \left(\begin{array}{cc} \rho & \epsilon \rho \\
                           0&0\end{array}\right), 
\label{minimo}
\ee
where $\epsilon= -b_{a0}/|b_{a0}|$ is the phase of ${\tilde h}_1$. 
The unbroken group $H$ corresponding to the minimum in (\ref{minimo}) 
is then generated by $\theta_1= \theta_2 =0$ and $\theta = \epsilon \theta_3$. 
The vacua associated to the coset space $G/H$ can be parametrized as 
\be
\Phi ({\bf \theta})= {\rm e}^{-i {\bf \theta} 
\cdot {\bf \sigma}} \Phi_0 {\rm e}^{- i \epsilon \theta_3 \sigma_1},
\label{gh}
\ee 
The three parameters in $\theta$ correspond to the three 
Goldstone bosons of the spontaneously broken $G$-symmetry. We will now derive 
the effective 
Lagrangian up to two derivatives describing them, once the bare mass terms 
and the mass spurions terms 
are 
introduced. These terms explicitly break the $SU(2)_{-} \times 
U(1)$ symmetry, and if they are small we can also compute exactly the induced 
masses for the Goldstone bosons \cite{weinberg}. 
The effective potential with the explicit breaking 
terms will have now a dependence on the $\theta$ angles, and we must find 
the minimum of $V_{\rm eff}$ in the $G/H$ space. We can write
\be
V_{{\rm eff}}= V_0 + {\rm Tr} \{{\bf M}^2 \Phi \Phi^{\dagger} \}- 
\sum_{i=1}^2 S_i^f \Phi^{ia} {\bf F}_f^{ab}( \Phi^{\dagger})^{bi},
\label{potencial}
\ee
where we have denoted by $V_0$ the $SU(2)_{-} \times U(1)$ symmetry-preserving 
terms, and the ${\Phi}$ matrix is the $\theta$-dependent one in (\ref{gh}). 
After some straightforward algebra we obtain:
\bea
V_{\rm eff}(\theta)&=&V_0+ 2 \rho ^2 \{ 2|a_1|^2- {\sqrt 2} 
\epsilon S_1^f {\rm Re}\,[F^f] \} \nonumber\\
&+&2\rho^2(1-{\hat  \theta_3}^2) {\rm sin}^2 |\theta| 
\{ 2(|a_2|^2-|a_1|^2)-{\sqrt 2}\epsilon {\rm Re}\,[F_{-}] \},
\label{vangle}
\eea
where $|\theta|=(\theta_1^2 + \theta_2^2 +\theta_3^2)^{1/2}$, and 
${\hat  \theta_i}= \theta_i/|\theta|$, $i=1,2,3$. There are two different 
minima, depending on 
the sign of the term multiplying $(1-{\hat  \theta_3}^2) {\rm sin}^2 |\theta|$. 

1) If $2(|a_2|^2-|a_1|^2)-{\sqrt 2}\epsilon {\rm Re}\,[F_{-}]>0$, the minimum 
is located at $\theta_1=\theta_2=0$ and $\theta_3$ arbitrary. We then see 
that one of the 
directions in $G/H$ remains flat, therefore one of the pions 
will remain massless. This is a consequence of the fact that
 the terms explicitly breaking $SU(2)_-$ in (\ref{potencial})
still leave an unbroken $U(1)_- \subset SU(2)_-$. 
If we consider small fluctuations around this minimum, 
it is natural to expand also around $\theta_3=0$, and we have then at first 
order
\be
 \Phi (\theta)=(1 - i{\bf \sigma}\cdot {\bf \xi})\Phi (\theta_{*}),
\label{expansion}
\ee
where $\theta_{*}=(0,0,0)$. The ${\xi}$-variables are given by
\be
\xi_1=\theta_1, \,\,\,\ \xi_2=\theta_2, \,\,\,\ \xi_3=2\theta_3.
\label{xis}
\ee
The potential (\ref{vangle}) can also be expanded to obtain
\bea
V_{\rm eff} &=& V_0 + 2 \rho^2 \{ |a_1|^2- {\sqrt 2} 
\epsilon S_1^f {\rm Re}\,[F^f] \} \nonumber\\
&+& 2\rho^2 \{ 2(|a_2|^2-|a_1|^2)-{\sqrt 2}\epsilon {\rm Re}\,[F_{-}] \}
(\xi_1^2 + \xi_2^2),
\label{vexp}
\eea
 To identify the 
pion fields ${\pi}_i$, $i=1,2,3$, we require the standard normalization 
for the kinetic terms,
\be
{\rm Tr} \{  {\partial_{\mu} } \Phi {\partial^{\mu} } {\Phi ^{\dagger}} \} = 
{1 \over 2} \sum_{i=1}^3  {\partial_{\mu} }\pi_i {\partial^{\mu} }\pi_i,
\label{kin}
\ee
which leads to 
\be
\pi_i = 2\rho  \xi_i, \,\,\,\,\,\ i=1,2,3.
\label{f}
\ee
 From (\ref{vexp}) and (\ref{f}) we finally obtain the pion mass and $F_{\pi}$ 
as
$$
M^2_{\pi_1} = M^2_{\pi_2}= 
 2(|a_2|^2-|a_1|^2)-{\sqrt 2}\epsilon {\rm Re}\,[F_{-}]
,\,\,\,\,\,\ M^2_{\pi_3}=0, 
$$
\be
F_{\pi}= 2 \rho,
\label{piones}
\ee
and we see that, indeed, one of the pions will remain massless.

2) If $2(|a_2|^2-|a_1|^2)-{\sqrt 2}\epsilon {\rm Re}\,[F_{-}]<0$, the minimum 
is located at $\theta_3=0$, $|\theta|=\pi/2$. We then parametrize 
\be
\theta_1=\Big( {\pi \over 2} + r \Big){\rm sin} \phi, \,\,\,\,\,\, 
\theta_2=\Big( {\pi \over 2} + r \Big){\rm cos} \phi,
\label{par}
\ee
so that the minimum occurs at $\theta_3=r=0$ and $\phi$ arbitrary. Again there 
is a flat direction and a massless pion. We consider small fluctuations around 
$(r, \phi, \theta_3)=(0,0,0)$, {\it i.e.}, $\theta_{*}=(0, \pi/2,0)$, and 
obtain
\be
\xi_1={2 \over \pi}\theta_3, \,\,\,\ \xi_2=r, \,\,\,\ \xi_3=-(\phi+\theta_3).
\label{xisdos}
\ee
Notice that $\Phi(\theta_{*})$ corresponds, for the minimum considered in (1), 
to a monopole VEV in the $h_1$ direction, and for the minimum considered here, 
to a VEV in the $h_2$ direction. These two different cases 
clearly show the anisotropy in the VEV direction once $SU(2)_{-}$ terms are 
introduced, as one direction or another is favoured depending on the 
value of $2(|a_2|^2-|a_1|^2)-{\sqrt 2}\epsilon {\rm Re}\,[F_{-}]$. As we are 
restricting the $\Phi$ values to $G/H$ configurations, the conditions 
to have one VEV direction or the other will only agree with the ones 
derived in section 4 in the case $F_f=0$ (as in this case the exact 
solution is in $G/H$). 
  
Once we have found the appropriate $\xi_i$ parameters, we can follow
 the procedure in (1) and we arrive to the same $F_{\pi}$ and to the same 
mass spectrum: one of the pions remains 
massless and for the other two we have the same result as in (\ref{piones})
but with the opposite sign (in such a way that the squared mass is always 
positive).

In both minima we have then the same physical situation, which we can 
summarize 
as:
$$
M^2_{\pi_1} = M^2_{\pi_2}= 
 |2(|a_2|^2-|a_1|^2)-{\sqrt 2}\epsilon {\rm Re}\,[F_{-}]|,\,\,\,\,\,\ 
M^2_{\pi_3}=0, 
$$
\be
F_{\pi}=2 \rho.
\label{todo}
\ee
The physical meaning of these equations is very appealing. They are telling us 
that, if chiral symmetry is broken by magnetic monopole condensation, 
then $F_{\pi}$ is given by the VEV of the monopole. Also notice that, for 
$F_{-}=0$, the pion mass is simply given by the difference between the 
BPS masses of the monopoles. If we turn on $F_{-}$, we can take into 
account that $\epsilon=-1$ and the structure of the minima discussed 
in section 4, to obtain the following expression for the mass of the 
pion $\pi = (\pi_1 + i \pi_2)/{\sqrt 2}$:
\be
M^2_{\pi}=2||a_2|^2-|a_1|^2|+{\sqrt 2}|{\rm Re}\,[F_{-}]|.
\label{finalmass}
\ee
Notice that the additional $SU(2)_-$ breaking  terms associated to $F_-$ 
give a positive contribution to the pion mass.
One can find an expansion for 
this expression in terms 
of $m_{\pm}$, $F_-$ and $f_0$ (as it has to be evaluated at the minima on the 
$u$-plane, whose position depends on these parameters). 
The first term of this expansion reads, for $F_-=0$, 
\be
M_{\pi}^2=cf_0|m_{-}||{\rm sin} \alpha | + O(m_f^2),
\label{masapion}
\ee
where $\alpha$ is the complex phase of $m_{-}$ and $c$ is an adimensional 
constant which depends on the variation of the minimum position w.r.t. $f_0$. 
We then see that, for real masses, the first non-zero term is 
{\it quadratic} in the 
bare quark masses. It reflects 
the fact that our softly broken model has squarks entering the 
Lagrangian with squared bare masses
and that the chiral symmetry breaking pattern is dominated by the 
squark condensates $<{\tilde q}_f q_f> \not=0$. 
In spite of the obvious differences, it 
is likely that some of these results for the pion Lagrangian will apply 
to more realisitic models.   

\newpage

{\large\bf Acknowledgements}

We acknowledge J.L.F. Barb\'on for a critical reading of the 
manuscript.
M.M. and F.Z. would like to thank the Theory 
Division at CERN
for its hospitality. 
The work of M.M.~is supported in part by DGICYT under grant
PB93-0344 and by
CICYT under grant AEN94-0928. The work of F.Z. is supported 
by a fellowship from Ministerio de Educaci\'on.

\end{document}